\documentclass[a4paper,11pt]{article}
\usepackage{jheppub}
\usepackage{amsmath,amssymb,multirow,array}
\usepackage{lingmacros}
\usepackage{amsmath,amssymb,multirow,array,graphicx,floatrow}
\usepackage{sidecap}
\usepackage{caption}
\usepackage{subcaption}
\usepackage[sort&compress,numbers]{natbib}
\usepackage{ifpdf}
\usepackage{setspace}
\usepackage{amsfonts}
\usepackage{color}
\usepackage{threeparttable}
\usepackage{graphicx}             
\DeclareGraphicsExtensions{.pdf}  
\usepackage{amsmath}              
\usepackage{feynmp}               

\usepackage{rotating}
\newcommand{\nc}{\newcommand}
\nc{\Gz}{\fullg}
\nc{\Gvc}{\boldsymbol{G}^c}
\nc{\Gam}{\boldsymbol{\Gamma}}
\nc{\rh}{\boldsymbol{\rho}}
\nc{\Sig}{\boldsymbol{\Sigma}}
\nc{\TS}{\tilde{\Sig}}
\nc{\TG}{\tilde{\mbox{\boldmath $G$}}}
\nc{\gam}{\boldsymbol{\gamma}}
\nc{\alp}{\boldsymbol{\alpha}}
\nc{\hs}{\hspace*{1mm}}
\nc{\fullg}{\boldsymbol{G}}
\nc{\scs}{\scriptstyle}
\nc{\beq}{\begin{eqnarray}}
\nc{\eeq}{\end{eqnarray}}
\nc{\la}{\label}
\nc{\no}{\nonumber}
\nc{\ci}{\cite}
\nc{\trace}{{\rm Tr\,}}
\nc{\setval}{\fmfset{wiggly_len}{1.5mm}\fmfset{arrow_len}{1.5mm}\fmfset{arrow_ang}{13}\fmfset{dash_len}{1.5mm}\fmfpen{0.125mm}\fmfset{dot_size}{1thick}}
\title{{ Monte Carlo Computation of Spectral Density Function in Real-Time Scalar Field Theory}}

\author{Navid Abbasi$^{a}$\footnote{Abbasi@ipm.ir}, \ Ali Davody$^a$\footnote{Davody@ipm.ir},\\
\small{\emph{$^{a}$ School of Particles and Accelerators, Institute for Research in Fundamental Sciences (IPM), P.O. Box 19395-5531, Tehran, Iran}} 
}

\abstract{
Non-perturbative study of "real-time" field theories is difficult due to the sign problem. 
We use Bold Schwinger-Dyson (SD) equations to study the real-time $\phi^4$ theory in $d=4$ beyond the perturbative regime. Combining SD equations in a particular way, we derive a non-linear integral equation for the two-point function. Then we introduce a new method by which one can analytically perform the momentum part of loop integrals in this equation. The price we must pay for such simplification is to numerically solve a non-linear integral equation for the spectral density function. Using Bold diagrammatic Monte Carlo method we find non-perturbative spectral function of theory and compare it with the one obtained from perturbation theory. At the end we utilize our Monte Carlo result to find the full vertex function as the basis for the computation of real-time scattering amplitudes.    }

\begin{document}
\maketitle

%
%
%


\section{Introduction}
Several decades after the field theory has been born,  lattice field theory is still regarded as
a powerful tool to study its non-perturbative aspects in the strong coupling regime. Lattice field theory 
is originally based
on computing the path integrals on Euclidean space accompanied by the stochastic sampling of partition function. Besides its successes however, it suffers from the well-known sign problem once it is supposed to apply to a system at finite chemical potential or to calculate the transport coefficients via the real time methods. 
"Finite size effect" is the other problematic characteristics of lattice field theory which is related to the limitations of numerics due to lattice's size. In quantum chromodynamics (QCD) for instance, this effect is an obstacle to investigate the theory in the strongly coupled infra-red regime as well.
   
In another direction from many years ago, people have tried to extract non-perturbative information from the Schwinger-Dyson (SD) equations \cite{Fischer:2006ub}.\footnote{Schwinger-Dyson equations are actually the equations of motion for the the Green's functions.} The common point among all these attempts has been to numerically solve SD equations on Euclidean space. Although it may seem special, but it has been enough to answer to some important problems like finding the beta function in QCD.    
On the other hand in order to compute the physical scattering amplitudes one must implement computations in the framework of real-time quantum field theory. However,  presence of unknown singular n-point functions on Lorentz space makes it so hard to do the regularization process numerically. 

Based on diagrammatic formulation of field theory, a new method has been developed to overcome the above-mentioned problems in recent years \cite{dmc2,hub}. The so-called Diagrammatic Monte Carlo (DMC) method performs a Monte Carlo process in the space of contributing Feynman diagrams using local Metropolis updates.
Unlike lattice filed theory, DMC is able to sample physical quantities in thermodynamics limit, however, due to divergence of 
perturbation series, a resummation technique is usually needed to make the scheme convergent. Specifically in \cite{Buividovich:2011zy}, Borel resummation technique has been used to establish the triviality of $\phi^4$ theory in four and five dimensions \cite{Luscher:1987ay}.
 
Towards improving convergence of DMC, the authors of \citep{bdmc1} have shown that representing of the physical quantities in some series of dressed (bold) correlators leads to a wider range of convergence compared to traditional way of expansion in terms of free correlators.  In their method, known as Bold Diagrammatic Monte Carlo (BDMC),
interestingly, the sign problem is not only an obstacle but also helps the scheme to better converge. Using BDMC, the equation of state of the "unitary Fermi gas"\footnote{A strongly interacting fermionic system.} at finite chemical potential has been studied recently \cite{nat}. 

Much more recently, one of us has used BDMC method in the context of relativistic quantum filed theory for the first time \cite{Davody:2013bua};  it has been shown that BDMC simply reproduces the fixed point value of renormalized coupling constant in a three dimensional $\phi^4$ theory living on Euclidean space. The result is in a very good agreement with the one obtained from lattice field theory. 
  Although there exist many other interesting problems in Euclidean quantum field theory, it seems more exciting to attack a real-time quantum field theory and to investigate its non-perturabtive aspects via using BDMC. As the first step in this way, we are going to non-perturabtively study the four dimensional $\phi^4$ theory through this note. Fundamentally, this theory is the underlying theory which describes the dynamics of Higgs boson in the standard model; however in the range of energies accessible in present experiments, it is essentially the IR limit of scalar theory which is of main importance for particle physicists. 
In contrast to scalar theory case,  quantum chromodynamics (QCD) becomes 
a strongly  coupled field theory in low energies and consequently out of control within the perturbation theory framework.  So, BDMC may behaves as a tool by which one may better explore the physics of low energy QCD.  This is exactly the final goal we want to achieve through a long project started from the present work. As it was pointed above, our starting point for applying BDMC method to real-time quantum field theory is 
to study $\phi^4$ theory as a toy example. Developing the method in this case might make it possible to extend the project to a more interesting theory as QCD. 

What actually we are going to do in this paper is to find the spectral density function of $\phi^4$ theory beyond the weak coupling regime.
Spectral density function is basically a non-perturbative object and cannot be captured through perturbation theory. To proceed non-perturbatively, we have to first derive real-time Schwinger-Dyson equations for full correlators of theory and express them in  terms of renormalized correlators as well. We shall show the full correlators including at most four field operators obey a pair of coupled\footnote{More precisely, these equations express both full two-point function $\textbf{G}$ and full vertex function $\Gam^{(4)}$ in terms of each other.} non-perturbative integral equations whose solutions determine both two-point and vertex function of theory. However, solving coupled integral equations accompanied with finding the renormalization coefficients is a hard task. A general way to unravel the problem is to truncate the vertex equation and to substitute it into the full equation of two-point function respectively. This results in a non-linear integral equation for the two-point function. The point which makes even numerically solving of the resultant equation complicated is the presence of unknown bold two-point functions under the momentum integrals. To simplify the numerics, we exploit the spectral representation of Green's function and rewrite the equation as an integral equation for the spectral density function.    
Surprisingly,  the momentum integrals in the latter equation can be simply performed through familiar method of perturbation theory. This process which we will refer to as the "Spectral Method" has a really valuable advantage: numerically performing the momentum integrals which is often problematic in Lorentz space will be completely replaced with analytically evaluating the quasi-perturbative loop integrals. Therefore our numerical problem reduces to just finding the spectral density function from a non-linear integral equation. It is the place where we will enter BDMC to solve the equation non-perturbatively.  It is worth noting that combining the spectral method with BDMC may greatly intensifies the efficiency of Monte Carlo in obtaining the spectral density function.

Having the spectral density function, one might be interested in finding the full vertex function as the basis for computation 
of scattering amplitudes in a typical process like $\phi \phi \rightarrow \phi \phi$. Similar to the two-point function case, SD equation of vertex function takes the form of an integral equation. However, since vertex function does not enjoy a spectral representation, we will introduce an alternative  scheme to find vertex function through a different way. To this end, we first develop using the well-known \textbf{Pad\'{e} Approximant of a function} to solve non-linear integral equations. Considering the SD equation of vertex function, we then find a convergent approximate solution for the vertex function.  In fact, spectral density function makes role of a non-perturbative link between approximate solution of vertex and our previous Monte Carlo computation.

In summary, our non-perturbative study of $\phi^4$ theory consists of two parts; firstly, to find spectral density function by making use of BDMC method and secondly, to get the result of first part in order to find an approximate Pad\'{e} solution for the vertex function. Let us emphasis that our main goal in this paper is just to introduce and to develop the above-mentioned two-part method.  It is why we do not proceed to make our results more accurate. It is obvious that the more number of terms kept in truncated vertex equation, the more accurate solutions we will obtain. We leave such rigorous computations to the case of  QCD in our future work. 

What we have provided in next sections is as follows; in section
(\ref{sec2})
we first derive Schwinger-Dyson equations of two-point and four-point function in the real-time $\phi^4$ theory. By introducing a set of renormalization coefficients, we then rewrite SD equations in terms of renormalized correlators. Finally we find two of coefficients via renormalization conditions and leave the last one undetermined.  In  section (\ref{sec3}), after representing the two point function in a bold expansion, we introduce the main part of the work, namely the spectral method, and then find the spectral density function via using BDMC method. In section (\ref{sec4}) we use the result obtained in previous section and investigate an approximate Pad\'{e} solution for the vertex function. Finally in section (\ref{sec5}) we end with pointing the open questions.

\section{Non-perturbative Formulation of Real-Time $\phi^4$ Theory }
\label{sec2}
In this section we are going to derive renormalized SD equations for real time $\phi^4$ theory, analogue of what has been already obtained for the four dimensional $\phi^4$ theory on Euclidean space .
In the following subsections we first derive real-time SD equations via some redefinitions in Euclidean SD equations of the same theory. As we will see, the resultant equations will express both two-point  and vertex functions as a pair of coupled recursive equations. Then,  we exchange every bare filed operator  with its renormalized version to find equations between renormalized correlators Finally we will determine two of three renormalization coefficients via relevant conditions, Let us start.
\footnote{We would like to thank to Farid Taghinavaz for collaborating in the first parts of this section.}

\subsection{To Derive Bare Real-Time Schwinger-Dyson Equations }
The real-time partition function of scalar theory may be formally given by  $Z=\int \mathcal{D}\phi\;e^{iS}$ with:\footnote{In what follows we always assume: $\eta^{\mu\nu}=\text{diag}(1,-1,-1,-1)$.
In addition, the subscript "b" always refers to the bare parameters while the subscript "r" denotes the renormalized parameters..}
\begin{equation}
\begin{split}
i\,S&= i\int d^dx\left[\frac{1}{2}\phi_{b}(x)\,(-\partial_\mu \partial^\mu-m_b^2)\,\phi_{b}(x)-\frac{g_b}{4!}\,\phi_{b}
^4(x)\,\right]\\
&= -\int d^dx\left[\frac{1}{2}\,\phi_{b}(x)\,i\,(\partial_\mu \partial^\mu+m_b^2)\,\phi_{b}(x)+i\,\frac{g_b}{4!}\,\phi_{b}^4(x)\,\right]
\end{split}
\end{equation}
As it is clear, in the second line we have rearranged terms and factors so that $iS$ looks like the Euclidean action $-S_{E}$.\footnote{One can impose the following replacement in Euclidean theory to simply derive the real-time results:
\beq \label{1.3}
G^{-1}_E(k^2)&\longrightarrow&  G^{-1}(k^2)= 
-i(k_\mu k^\mu-m_b^2+i \epsilon)\\ \label{1.4}
g_b&\longrightarrow& i\,g_b 
\eeq
where $i\epsilon$ prescription has been also considered.}
As a result one can derive SD equations for self-energy (or equivalently for two-point function) and vertex function just by applying (\ref{1.3}) and (\ref{1.4}) to Euclidean SD equations \cite{Pelster:2003rc}.  
Doing so,
 the self-energy takes the following form:\footnote{Let us emphasise that both
 propagator and vertex function are explicit function of Lorentz scalars, however for sake of simplicity we sometimes write the vertex function as a function of its external momenta.}
\begin{equation}\label{1.6}
\begin{split}
\Sig_{b}(p^2)&\equiv\Gamma^{(2)}_b(p^2)-\Gam^{(2)}_b(p^2)\\
&=- i\, \frac{g_b}{2} \int\frac{d^dk}{(2\pi)^d}\, \Gz_b(k^2)
-i\, \frac{g_b}{6} \int\frac{d^dk}{(2\pi)^d}\,\frac{d^dq}{(2\pi)^d}\,\Gz_{b}(k^2)\; \Gz_{b}(q^2) \Gz_{b}(l^2)
\Gam^{(4)}_{b}(p,k,q,l)
\end{split}
\end{equation}
where $l^{\mu}=p^{\mu}-k^{\mu}-q^{\mu}$. In this equation and all next expressions, $\Gam^{(2)}(p^2)$ refers to the inverse  propagator.
Note that the first term is obviously a one-loop integral while the second one represents a two-loop contribution. By one- or two-loop here, we do not mean the familiar perturbation theory loops; these are really non-perturbative loops including dressed propagators; we will discuss more on this issue in next section. In addition to integral form, it is possible to graphically represent equation (\ref{1.6}) as the following:

\begin{equation}\label{graphself}
(\;\parbox{7mm}{\begin{fmffile}{Feynman/e1}
 	\begin{fmfgraph*}(15,10)
	   \fmfleft{i} \fmfright{o}

\fmf{plain,tension=.8}{i,v,o}
\fmfv{decor.shape=circle,decor.filled=30,decor.size=.33w}{v}
	\end{fmfgraph*}
\end{fmffile}
})^{-1}
=
(\;\parbox{7mm}{\begin{fmffile}{Feynman/e2}
 	\begin{fmfgraph*}(15,10)
	   \fmfleft{i} \fmfright{o}
\fmf{plain,tension=1}{i,v,o}
	\end{fmfgraph*}
\end{fmffile}
})^{-1}
+{\scalebox{1}{$\frac{1}{2}$}}
 \parbox{15mm}{\begin{fmffile}{Feynman/e3}
 	\begin{fmfgraph*}(40,30)
	   \fmfleft{i1}
	   \fmfright{o1}
		\fmftop{d2}
	\fmf{plain}{i1,d1,o1}
		\fmffreeze
	\fmf{plain,left}{d1,d2,d1}
\fmfv{decor.shape=circle,decor.filled=30,decor.size=.13w}{d2}
	\fmfshift{(.2w,0)}{i1}
		\fmfshift{(-.2w,0)}{o1}
	\end{fmfgraph*}
\end{fmffile}
}
+{\scalebox{1}{$\frac{1}{6}$}}\;
 \parbox{25mm}{\begin{fmffile}{Feynman/e4}
 		\begin{fmfgraph}(40,40)
		\fmfleft{i}
		\fmfright{o}
	\fmftop{t}
	\fmfbottom{b}
\fmf{plain,tension=5}{i,v1,v2,v3,o}		
\fmffreeze	
	\fmf{plain,left,tension=0.4}{v1,v3}
	\fmf{plain,right,tension=0.4}{v1,v3}
		\fmfshift{(0,-.25h)}{t}
		\fmfshift{(0,.25h)}{b}
		\fmfv{decor.shape=circle,decor.filled=full,decor.size=.15w}{v3}
		\fmfv{decor.shape=circle,decor.filled=30,decor.size=.12w}{v2,t,b}

	\end{fmfgraph}
\end{fmffile}
}
\end{equation}
where each propagator with a gray dot represents a dressed (full) propagator and the full four-point (vertex) function has been pictured by a black dot.
Clearly,  (\ref{1.6}) and (\ref{graphself}) are exact equations including both perturbative and non perturbative
information about the theory.

Similarly, the SD of vertex function can be derived from its Euclidean version. After some lengthy but straightforward calculations one reaches to:
\beq\label{1.7}
\Gam^{(4)}_{b}(p_{1},p_{2},p_{3},p_{4}) = ig_{0}\left(1+\mathcal{A}_b(p_{i}.p_{j})+ \mathcal{B}_b(p_{i}.p_{j})+
\frac{}{} \mathcal{D}_b(p_{i}.p_{j})+\overline{\mathcal{D}}_b(p_{i}.p_{j})\right),
\eeq
where $i,j \in \{1,2,3,4\}$ and 
\begin{equation}\label{1.8}
\begin{split}
\mathcal{A}_b(p_{i}.p_{j}) & = -\frac{1}{2}\; \underbrace{\Gz_{b}(p_{5}^2)\,
 \Gz_{b}(p_{6}^2)\,  \Gam^{(4)}_{b}(p_{5},p_{6},p_{3},p_{4})}_{p_{6}^{\mu}=p_{1}^{\mu}+p_{2}^{\mu}-p_{5}^{\mu}}\,+\,(2\leftrightarrow 3,4)\\
 \mathcal{B}_b(p_{i}.p_{j}) & =\frac{1}{6}
 \;\underbrace{\Gz_{b}(p_{5}^2)\; \Gz_{b}(p_{6}^2)\,
  \Gz_{b}(p_{7}^2)\,  \Gz_{b}(p_{8}^2)\, \Gam^{(4)}_{b}(p_{6},p_{7},p_{2},-p_{8})\;\Gam^{(4)}_{b}(p_{8},p_{5},p_{3},p_{4})}_{p^{\mu}_{8}=p^{\mu}_{1}+p^{\mu}_{2}-p^{\mu}_{5},\,\,\,\,\,p^{\mu}_{7}=p^{\mu}_{1}-p^{\mu}_{6}-p^{\mu}_{5}}+(2\leftrightarrow 3,4)\\
\mathcal{D}_{b}(p_{i}.p_{j}) & =  -\frac{1}{3}\; \Gz_{b}(p_{5}^2)\,
 \Gz_{b}(p_{6}^2)\,  \Gz_{b}(p_{7}^2)\, \frac{\delta \Gam^{(4)}_{b}(p_{5},p_{2},p_{3},p_{4})}{\delta \Gz_{b}(-p_{6},-p_{7})}\\
 \overline{\mathcal{D}}_{b}(p_{i}.p_{j}) & =  \frac{1}{6}\;\Gz_{b}(p_{5}^2)\,
  \Gz_{b}(p_{6}^2)\,  \Gz_{b}(p_{7}^2)\,\Gz_{b}(p_{9}^2)\,  \Gz_{b}(p_{0}^2)
    \Gam^{(4)}_{b}( p_{6},p_{7},p_{9},p_{0}) \frac{\delta \Gam^{(4)}_{b}(p_{5},p_{2},p_{3},p_{4})}{\delta \Gz_{b}(p_{9},p_{0})}
\end{split}
\end{equation}
There are some points about these equations  which need to be clarified. First, for sake of brevity, we have dropped all loop integrals in writing above expressions; however note that the momentum conservation condition, $p^{\mu}_{1}+p^{\mu}_{2}+p^{\mu}_{3}+p^{\mu}_{4}=0$, has to be always considered.  Second, in the expressions $\mathcal{A}_b$ and $\mathcal{B}_b$, "$(2\leftrightarrow 3,4)$" denotes that there are another two terms similar to the underbraced term which  can be simply written through exchanging $p_{2}$ with $p_{3}$ and $p_{2}$ with $p_{4}$.  Lastly, in each expression, by use of conservation of the momentum at vertices, we have written  non independent internal momenta in terms of independent ones, under a brace. In the case of two last expressions that contains functional derivative factors, we will discuss in detail in (\ref{AppDer}).

As it is given above, vertex function in $\phi^{4}$ theory is a complicated object including many terms with several permutations.
Similar to the case of inverse propagator, we may rewrite such long expressions in a more concise representation, namely graphical SD equation, as it follows:

\begin{eqnarray}\label{graphvertex}
\parbox{8mm}{\begin{fmffile}{Feynman/v0}
		\begin{fmfgraph*}(15,15)
		\fmfleft{i1,i2}
		\fmfright{o1,o2}

		\fmf{plain}{i1,v1,o2}
		\fmf{plain}{i2,v2,o1}

\fmfv{decor.shape=circle,decor.filled=full,decor.size=.35w}{v2}

	\fmfv{l=${\scriptstyle 1 }$,l.a=-120,l.d=.1w}{i1}
	\fmfv{l=${\scriptstyle 2 }$,l.a=120,l.d=.1w}{i2}
	\fmfv{l=${\scriptstyle 4 }$,l.d=.1w}{o1}
	\fmfv{l=${\scriptstyle 3 }$,l.d=.1w}{o2}

		\end{fmfgraph*}
	\end{fmffile}
}
&=&\;	
\parbox{8mm}{\begin{fmffile}{Feynman/v1}
	\begin{fmfgraph*}(15,15)
	\fmfleft{i1,i2}
	\fmfright{o1,o2}
	\fmf{plain}{i1,v1,o2}
	\fmf{plain}{i2,v2,o1}
	
		\fmfv{l=${\scriptstyle 1 }$,l.a=-120,l.d=.1w}{i1}
		\fmfv{l=${\scriptstyle 2 }$,l.a=120,l.d=.1w}{i2}
		\fmfv{l=${\scriptstyle 4 }$,l.d=.1w}{o1}
		\fmfv{l=${\scriptstyle 3 }$,l.d=.1w}{o2}
	\end{fmfgraph*}
	\end{fmffile}
}
-{\scalebox{1}{$\frac{1}{2}$}}\;\;
\parbox{19mm}{\begin{fmffile}{Feynman/v2}
	\begin{fmfgraph*}(50,30)
	\fmfleft{i1,i2}
	\fmfright{o1,o2}
	\fmfv{l=${\scriptstyle 1}$,l.a=-120,l.d=.05w}{i1}
	\fmfv{l=${\scriptstyle 2 }$,l.a=120,l.d=.05w}{i2}
	\fmfv{l=${\scriptstyle 4 }$,l.a=-40,l.d=.05w}{o1}
	\fmfv{l=${\scriptstyle 3 }$,l.a=60,l.d=.05w}{o2}
	\fmftop{t}
	\fmfbottom{b}
	\fmf{plain,tension=5}{i1,v1,i2}
	\fmf{plain,tension=5}{o1,v2,o2}
	\fmf{plain,left=.5,tension=.6}{v1,v2}	
	\fmf{plain,right=.5,tension=.6}{v1,v2}
	\fmfv{decor.shape=circle,decor.filled=full,decor.size=.12w}{v2}
	\fmfv{decor.shape=circle,decor.filled=30,decor.size=.1w}{t}
	\fmfv{decor.shape=circle,decor.filled=30,decor.size=.1w}{b}
	\fmfshift{(0,-0.25h)}{t}
	\fmfshift{(0,0.25h)}{b}
	\fmfshift{( 0.,-0.35h)}{i2}
	\fmfshift{( 0.,0.35h)}{i1}
	
	\fmfshift{( 0.,-0.3h)}{o2}
	\fmfshift{( 0.,0.3h)}{o1}

	\end{fmfgraph*}
	\end{fmffile}
}-{\scalebox{1}{$\frac{1}{2}$}}\;\;
\parbox{19mm}{\begin{fmffile}{Feynman/v2p1}
	\begin{fmfgraph*}(50,30)
	\fmfleft{i1,i2}
	\fmfright{o1,o2}
	\fmfv{l=${\scriptstyle 1}$,l.a=-120,l.d=.05w}{i1}
	\fmfv{l=${\scriptstyle 3 }$,l.a=120,l.d=.05w}{i2}
	\fmfv{l=${\scriptstyle 4 }$,l.a=-40,l.d=.05w}{o1}
	\fmfv{l=${\scriptstyle 2 }$,l.a=60,l.d=.05w}{o2}
	\fmftop{t}
	\fmfbottom{b}
	\fmf{plain,tension=5}{i1,v1,i2}
	\fmf{plain,tension=5}{o1,v2,o2}
	\fmf{plain,left=.5,tension=.6}{v1,v2}	
	\fmf{plain,right=.5,tension=.6}{v1,v2}
	\fmfv{decor.shape=circle,decor.filled=full,decor.size=.12w}{v2}
	\fmfv{decor.shape=circle,decor.filled=30,decor.size=.1w}{t}
	\fmfv{decor.shape=circle,decor.filled=30,decor.size=.1w}{b}
	\fmfshift{(0,-0.25h)}{t}
	\fmfshift{(0,0.25h)}{b}
	\fmfshift{( 0.,-0.35h)}{i2}
	\fmfshift{( 0.,0.35h)}{i1}
	
	\fmfshift{( 0.,-0.3h)}{o2}
	\fmfshift{( 0.,0.3h)}{o1}

	\end{fmfgraph*}
	\end{fmffile}
}
-{\scalebox{1}{$\frac{1}{2}$}}\;\;
\parbox{19mm}{\begin{fmffile}{Feynman/v2p2}
	\begin{fmfgraph*}(50,30)
	\fmfleft{i1,i2}
	\fmfright{o1,o2}
	\fmfv{l=${\scriptstyle 1}$,l.a=-120,l.d=.05w}{i1}
	\fmfv{l=${\scriptstyle 4 }$,l.a=120,l.d=.05w}{i2}
	\fmfv{l=${\scriptstyle 3 }$,l.a=-40,l.d=.05w}{o1}
	\fmfv{l=${\scriptstyle 2 }$,l.a=60,l.d=.05w}{o2}
	\fmftop{t}
	\fmfbottom{b}
	\fmf{plain,tension=5}{i1,v1,i2}
	\fmf{plain,tension=5}{o1,v2,o2}
	\fmf{plain,left=.5,tension=.6}{v1,v2}	
	\fmf{plain,right=.5,tension=.6}{v1,v2}
	\fmfv{decor.shape=circle,decor.filled=full,decor.size=.12w}{v2}
	\fmfv{decor.shape=circle,decor.filled=30,decor.size=.1w}{t}
	\fmfv{decor.shape=circle,decor.filled=30,decor.size=.1w}{b}
	\fmfshift{(0,-0.25h)}{t}
	\fmfshift{(0,0.25h)}{b}
	\fmfshift{( 0.,-0.35h)}{i2}
	\fmfshift{( 0.,0.35h)}{i1}
	
	\fmfshift{( 0.,-0.3h)}{o2}
	\fmfshift{( 0.,0.3h)}{o1}

	\end{fmfgraph*}
	\end{fmffile}
}+\;{\scalebox{1}{$\frac{1}{6}$}}\;
\parbox{13mm}{\begin{fmffile}{Feynman/v3}
	\begin{fmfgraph*}(30,25)
		\fmfsurroundn{v}{12} 
		\fmfshift{(0,0.2h)}{v3}
		\fmfshift{(0,0.2h)}{v5}
		\fmfshift{(0,0.0h)}{v4}
		\fmfshift{(-.3w,0.0h)}{v9}
		\fmfshift{(.3w,0.0h)}{v11}
		\fmfshift{(-.5w,-0.01h)}{v1}
		\fmfshift{(0.w,-0.02h)}{v10}
		\fmfshift{(.22w,0.2h)}{v7}
		\fmfshift{(-.18w,-0.08h)}{v2}
		\fmf{plain,tension=1}{v5,v4,v3}
		\fmf{plain,right=.2,tension=1}{v4,v8}
		\fmf{plain,left=.2,tension=1}{v4,v12}
		\fmf{plain,left=.5,tension=1}{v8,v12}
		\fmf{plain,right=.5,tension=1}{v8,v12}
		\fmf{plain,left=.1,tension=1}{v9,v8}
		\fmf{plain,right=.1,tension=1}{v11,v12}
		\fmfv{decor.shape=circle,decor.filled=full,decor.size=.18w}{v4,v8}
		\fmfv{decor.shape=circle,decor.filled=30,decor.size=.16w}{v10,v1,v7,v2}
	\fmfv{l=${\scriptstyle 3}$,l.a=30,l.d=.05w}{v3}
	\fmfv{l=${\scriptstyle 2 }$,l.a=120,l.d=.05w}{v5}
\fmfv{l=${\scriptstyle 1 }$,l.a=-120,l.d=.05w}{v9}
\fmfv{l=${\scriptstyle 4 }$,l.a=-30,l.d=.05w}{v11}
	\end{fmfgraph*}
	\end{fmffile}
}
\nonumber 
\\[5mm]
&&\;\;\;\;\;\;+\;{\scalebox{1}{$\frac{1}{6}$}}\;\;
\parbox{13mm}{\begin{fmffile}{Feynman/v3p1}
	\begin{fmfgraph*}(30,25)
		\fmfsurroundn{v}{12} 
		\fmfshift{(0,0.2h)}{v3}
		\fmfshift{(0,0.2h)}{v5}
		\fmfshift{(0,0.0h)}{v4}
		\fmfshift{(-.3w,0.0h)}{v9}
		\fmfshift{(.3w,0.0h)}{v11}
		\fmfshift{(-.5w,-0.01h)}{v1}
		\fmfshift{(0.w,-0.02h)}{v10}
		\fmfshift{(.22w,0.2h)}{v7}
		\fmfshift{(-.18w,-0.08h)}{v2}
		\fmf{plain,tension=1}{v5,v4,v3}
		\fmf{plain,right=.2,tension=1}{v4,v8}
		\fmf{plain,left=.2,tension=1}{v4,v12}
		\fmf{plain,left=.5,tension=1}{v8,v12}
		\fmf{plain,right=.5,tension=1}{v8,v12}
		\fmf{plain,left=.1,tension=1}{v9,v8}
		\fmf{plain,right=.1,tension=1}{v11,v12}
		\fmfv{decor.shape=circle,decor.filled=full,decor.size=.18w}{v4,v8}
		\fmfv{decor.shape=circle,decor.filled=30,decor.size=.16w}{v10,v1,v7,v2}
	\fmfv{l=${\scriptstyle 2}$,l.a=30,l.d=.05w}{v3}
	\fmfv{l=${\scriptstyle 3 }$,l.a=120,l.d=.05w}{v5}
\fmfv{l=${\scriptstyle 1 }$,l.a=-120,l.d=.05w}{v9}
\fmfv{l=${\scriptstyle 4 }$,l.a=-30,l.d=.05w}{v11}
	\end{fmfgraph*}
	\end{fmffile}
}
+
{\scalebox{1}{$\frac{1}{6}$}}\;\;
\parbox{13mm}{\begin{fmffile}{Feynman/v3p2}
	\begin{fmfgraph*}(30,25)
		\fmfsurroundn{v}{12} 
		\fmfshift{(0,0.2h)}{v3}
		\fmfshift{(0,0.2h)}{v5}
		\fmfshift{(0,0.0h)}{v4}
		\fmfshift{(-.3w,0.0h)}{v9}
		\fmfshift{(.3w,0.0h)}{v11}
		\fmfshift{(-.5w,-0.01h)}{v1}
		\fmfshift{(0.w,-0.02h)}{v10}
		\fmfshift{(.22w,0.2h)}{v7}
		\fmfshift{(-.18w,-0.08h)}{v2}
		\fmf{plain,tension=1}{v5,v4,v3}
		\fmf{plain,right=.2,tension=1}{v4,v8}
		\fmf{plain,left=.2,tension=1}{v4,v12}
		\fmf{plain,left=.5,tension=1}{v8,v12}
		\fmf{plain,right=.5,tension=1}{v8,v12}
		\fmf{plain,left=.1,tension=1}{v9,v8}
		\fmf{plain,right=.1,tension=1}{v11,v12}
		\fmfv{decor.shape=circle,decor.filled=full,decor.size=.18w}{v4,v8}
		\fmfv{decor.shape=circle,decor.filled=30,decor.size=.16w}{v10,v1,v7,v2}
	\fmfv{l=${\scriptstyle 2}$,l.a=30,l.d=.05w}{v3}
	\fmfv{l=${\scriptstyle 4 }$,l.a=120,l.d=.05w}{v5}
\fmfv{l=${\scriptstyle 1 }$,l.a=-120,l.d=.05w}{v9}
\fmfv{l=${\scriptstyle 3 }$,l.a=-30,l.d=.05w}{v11}
	\end{fmfgraph*}
	\end{fmffile}
}-\;
{\scalebox{1}{$\frac{1}{3}$}}\;\;\;\;
\parbox{13mm}{\begin{fmffile}{Feynman/v4}
		\begin{fmfgraph*}(37,37)
\fmfsurroundn{v}{12} 

\fmfv{decor.shape=hexagon,decor.filled=empty,decoration.angle=90,
decor.size=.4w}{v1}

	\fmfshift{(.2w,-0.h)}{v12}
	\fmfshift{(0.2w,0.0h)}{v2}
	\fmf{plain,tension=1}{v12,v1}
	\fmf{plain,tension=1}{v1,v2}
	\fmf{plain,right=.7,tension=1}{v1,v7}
	\fmf{plain,left=.7,tension=1}{v1,v7}
	\fmf{plain,tension=1}{v1,v7}
	
	\fmfshift{(-0.2w,-0.265h)}{v6}
	\fmf{plain,tension=1}{v7,v6}

		\fmfshift{(.55w,.45h)}{v11}
		\fmf{plain,tension=1}{v11,v1}

	\fmfshift{(-0.02w,-0.16h)}{v4}
	
	\fmfshift{(-0.03w,0.16h)}{v10}
	
	\fmfshift{(0.25w,0.45h)}{v9}

\fmfv{decor.shape=circle,decor.filled=30,decor.size=.16w}{v4,v10,v9}

\fmfv{l=${\scriptstyle 3}$,l.a=0,l.d=.05w}{v11}
\fmfv{l=${\scriptstyle 2 }$,l.a=60,l.d=.05w}{v2}
\fmfv{l=${\scriptstyle 4 }$,l.a=-60,l.d=.05w}{v12}
\fmfv{l=${\scriptstyle 1 }$,l.a=180,l.d=.05w}{v6}

	\end{fmfgraph*}

	\end{fmffile}
}\;\;\;\;\;\;
+{\scalebox{1}{$\frac{1}{6}$}}\;\;\;\;\;
\parbox{18mm}{\begin{fmffile}{Feynman/v5}
			\begin{fmfgraph*}(40,40)
\fmfsurroundn{v}{12} 

\fmfv{decor.shape=hexagon,decor.filled=empty,decoration.angle=90,
decor.size=.4w}{v1}

	\fmfshift{(.2w,-0.h)}{v12}
	\fmfshift{(0.2w,0.0h)}{v2}
	\fmf{plain,tension=1}{v12,v1}
	\fmf{plain,tension=1}{v1,v2}
	
	\fmf{plain,right=.7,tension=1}{v1,v7}

	\fmf{plain,left=.37,tension=1}{v1,v10}
	\fmf{plain,right=.4,tension=1}{v1,v10}
	
\fmf{plain,left=.37,tension=1}{v10,v7}
\fmf{plain,right=.4,tension=1}{v10,v7}

	\fmfv{decor.shape=circle,decor.filled=full,decor.size=.15w}{v10}	
	2
	\fmfshift{(-0.21w,-0.27h)}{v6}
	\fmf{plain,tension=1}{v7,v6}

		\fmfshift{(.55w,.45h)}{v11}
		\fmf{plain,tension=1}{v11,v1}

	\fmfshift{(-0.02w,-0.16h)}{v4}
	
	\fmfshift{(-0.03w,0.16h)}{v10}
	
	\fmfshift{(0.25w,0.45h)}{v9}

	\fmfv{decor.shape=circle,decor.filled=30,decor.size=.15w}{v3,v5,v4,v9,v8}

	\fmfshift{(-0.12w,-0.52h)}{v3}
	
		\fmfshift{(0.25w,-0.3h)}{v9}

		\fmfshift{(0.05w,-0.52h)}{v5}
	
			\fmfshift{(0.15w,0.01h)}{v8}
	\fmfv{l=${\scriptstyle 3}$,l.a=0,l.d=.05w}{v11}
	\fmfv{l=${\scriptstyle 2 }$,l.a=60,l.d=.05w}{v2}
	\fmfv{l=${\scriptstyle 4 }$,l.a=-60,l.d=.05w}{v12}
	\fmfv{l=${\scriptstyle 1 }$,l.a=180,l.d=.05w}{v6}
	
	\end{fmfgraph*}
	
		\end{fmffile}}	
\end{eqnarray}


Before ending this subsection let us give a few comments in order about the graphical equations (\ref{graphself}) and (\ref{graphvertex}). Firstly, presence of purely one-particle-irreducible graphs in RHS of these equations leads to not happening any cancellation between different terms through renormalization process, something that potentially intensifies  efficiency of our Monte Carlo computations \cite{Davody:2013bua}. Secondly, it is worth mentioning that such  representation for SD equations was already used in \citep{Pelster:2003rc} for the case of scalar field theory on Euclidean space. The point which distinguishes our equations from those in \citep{Pelster:2003rc} is that while in equation (100) in \citep{Pelster:2003rc} there is a derivative term containing internal bare lines, our equations are fully represented in terms of dressed (bold) internal lines.

\subsection{Schwinger-Dyson Equations with Renormalized  Operators }
In order to start renormalizing the theory, we need to introduce the renormalization coefficients.
Following the usual text book definitions, we may rewrite the action as:
\begin{equation}  \label{Sr1}
i\,S  =- \int d^d x \left( \frac{1}{2} \,
\phi (x)\,i\, (Z \partial^2 +Z_m m^2) \, \phi (x) + i\,\frac{g}{4!}Z_g
 \phi^4 (x)  \right)
 \end{equation}
where the set of three renormalization coefficients $\{Z,Z_m,Z_g\}$ have been used to associate the renormalized and bare (with subscript b) parameters to each other.
One might have been noticed that in writing (\ref{Sr1}), we have arranged different terms in a way so that the action seems similar to the bare action. 
Having the action in this form allows us to simply derive the renormalized version of SD equations just by imposing the following replacements in bare SD equations, namely in (\ref{1.6}) and (\ref{1.7}):
\beq
\begin{split}   \label{1.21}
\Gz_b&\longrightarrow \Gz\\ \label{1.22}
\Gam^{(4)}_b&\longrightarrow\Gam^{(4)}\\
G_b^{-1}(p^2)=i(-p^2+m_b^2)&\longrightarrow
i(-\,Z\,p^2+Z_m\,m^2)\\ \label{1.223}
g_b&\longrightarrow Z_g\, g.
\end{split}
\eeq
As a result, we find the equation of renormalized inverse propagator as the following: 
\beq\label{1.14}
\Gam^{(2)}(p^2)=-i\,\left( Z\,p^2-Z_m\,m^2+i\frac{}{}\epsilon+\,Z_g\,\Sig(p^2)\right)
\eeq
with $\Sig(p^2)$ defined the same as (\ref{1.6}) but with dropping the subscripts and applying (\ref{1.223}).
 Relatedly, the vertex function is given by ($i,j\in\{1,2,3,4\}$):
\beq\label{1.15}
\Gam^{(4)}(p_{1},p_{2},p_{3},p_{4})&  =  i\, Z_g\,g\left(1+\,\mathcal{A}(p_{i}.p_{j})+\,\mathcal{B}(p_{i}.p_{j})+ \,\mathcal{D}(p_{i}.p_{j})+\,\overline{\mathcal{D}}(p_{i}.p_{j})\right),\eeq
where the functions $\mathcal{A}$, $\mathcal{B}$, $\mathcal{D}$ and 
$ \overline{\mathcal{D}}$ are expectedly defined via the same definitions given in (\ref{1.8}) after dropping the subscripts "b".
It has not to be forgotten that $\Gam^{(4)}$ is not an explicit function of four momenta of its four legs. In general, it is a function of six Lorentz scalars made out of those momenta.

So for, we have derived SD equations for two- and four-point functions with undetermined 
renormalization coefficients. In next subsection we introduce a set of relevant conditions by which one would be able to find these coefficients.

\subsection{To Find Renormalization Coefficients}
SD equations obtained in previous subsection include three undetermined renormalization coefficients.  In order to fix the value of these coefficients, the same number of renormalization conditions is needed. So one may formally write: 
\begin{eqnarray}
\begin{split}
\la{rc11}
\Gam^{(2)}\,(p^2=m^2)\,\,\,&=\,\,\epsilon\\  \la{rc22}
\frac{\partial}{\partial p^2}\Gam^{(2)}\,(p^2)\vert_{p^2=m^2}\,\,\,&=\,\,-i\\
\la{rc33}
\Gam^{(4)}(s=4m^2,t=0,u=0)\,\,\,&=\,\,i\,g.
\end{split}
\end{eqnarray}
According to our previously made notation for $\Gam^{(4)}$, it is also possible to rewrite the third condition of (\ref{rc11}) as 
$\Gam^{(4)}(p,p,-p,-p)=ig$
with:
\begin{equation}\label{1.20}
p^{\mu}_{1}=p^{\mu}_{2}=(m,0,0,0)\equiv p,\,\,\,\,\,\,\,\,\,\,p^{\mu}_{3}=p^{\mu}_{4}=(-m,0,0,0)\equiv -p.
\end{equation}
 Imposing the first two conditions of (\ref{rc22}), one can simply find the coefficients $Z$ and $Z_{m}$ in terms of $Z_{g}$.
As a result, (\ref{1.14}) can be given as the following:
\beq\label{1.29}
\Gam^{(2)}(p^2)=-i\,(p^2-m^2+i\,\epsilon) -i\,\,Z_g\,\left(\Sig(p^2)-\Sig(m^2)-\frac{d\Sig(p^2)}{dp^2}|_{p^2=m^2}(p^2-m^2)\right).
\eeq
We do not intend to determine the last unknown coefficient, $Z_{g}$, here. We leave its computation to next  sections  wherever we will enter the full-propagator expansion into our computations. 
Correspondingly, we do not make any effort now to more simplify the second SD equation, namely equation (\ref{1.15}).

To summarize, we have derived the set of SD equations containing dressed two- and four-point functions in the real-time scalar field theory. These equations are a pair of coupled integral equations giving by (\ref{1.15}) and (\ref{1.29}). Although it would be possible to derive higher order SD equations, we limit our following discussions to specifically  the two above-mentioned equations.

\section{Bold Expansion: Dressed Propagator in Terms of Dressed Propagator}\label{sec3}
 Extracting information from the coupled non-linear equations derived in last section is so hard, even numerically. Furthermore, as we pointed out in the introduction, it has been shown that the bold expansion is usually converges more quickly than the well known perturbative expansion on the coupling constant.
 So in order to non-perturbatively study the theory,  we start to
rearrange each of SD equations in a series of "\textbf{renormalized full propagator}" i.e. $\textbf{G}$; this is what we refer to as the \textbf{bold expansion}.
We will devote the whole of this section to study the non-perturbative aspects of two-point function through solving its bold SD equation.
\subsection{Analytically Performing Dressed Loop Integrals: Spectral Density Function Method}
 From now on, rather than working with inverse propagator (\ref{1.29}), we prefer to have the bold expansion of dressed propagator 
itself. To this end we multiply equation (\ref{1.29}) by $G(p^2)\textbf{G}(p^2)$ and obtain 
\begin{equation}\label{3.1}
\textbf{G}(p^2)=G(p^2)+G(p^2)\textbf{G}(p^2)\textbf{Y}(p^2)
\end{equation}
where, via using (\ref{1.6}), $\textbf{Y}(p^2)$ turns out to be as:
\begin{equation}\label{3.2}
\textbf{Y}(p^2)=i Z_{g}\left(\Sig(p^2)-\Sig(m^2)-\frac{d\Sig(p^2)}{dp^2}\vert_{p^2=m^2}(p^2-m^2)\right).
\end{equation}
As it was already indicated, our goal in this paper is exploiting the power of bold expansion to find field theoretic results
beyond the weak coupling regime. In this way, we restrict our computations in the current section to the lowest order of approximation and leave more precise computations to our next work.  By the lowest order, we mean that each vertex function in RHS of (\ref{3.1}) is approximated by its tree level value, $ig$. This is in fact the leading contribution in truncated SD equation of vertex function. 
Neglecting  the first term, (\ref{1.6}) may be written as the following\footnote{Note that the 
first term in (\ref{1.6}) (and of course in its renormalized version) is independent of $p^2$ and consequently does not contribute to $\textbf{Y}(p^2)$.}:
\begin{equation}\label{3.3}
\Sig(p^2)=\frac{i g^2}{6}\int \frac{d^4 k d^4 q}{(2 \pi)^8} \textbf{G}(k^2)\textbf{G}(q^2)\textbf{G}(l^2)
\end{equation}
where $l^{\mu}=p^{\mu}-k^{\mu}-q^{\mu}$. 
Equation (\ref{3.3}), in RHS, contains a two-loop integral which is basically divergent. 
Graphically, this integral would correspond to the well-known "sunset" Feynman diagram if the propagators were not dressed.

The presence of full propagators in (\ref{3.3})
 makes the integral impossible to be performed. It is because the analytic form of the dressed propagators is unknown. 
If we even knew the analytic expression of bold propagators, the worse problem would be how to numerically evaluate
the integrals in Lorentz space, since no one could guaranty that the integrals could be analytically continued to Euclidean space.\footnote{Let us remind that in perturbation quantum field theory we always deal with free propagators in the integrands which can be simply transformed into non singular expressions in Euclidean space through Wick rotation.}  
For this reason we introduce and develop a new interesting method in order to transform the integrals to a new form which could be better under the control. The idea is so simple; to  work with the \textbf{"spectral density function"}, $\rho({p^2})$. In Quantum field theory, spectral density function is often defined in the context of Callan-Lehman representation as the following \citep{Schartz}
\begin{equation}\label{2.12}
\textbf{G}(k^2)=i \int d \mu^2 \, \frac{\rh(\mu^2)}{k^2- \mu^2+i \epsilon}.
\end{equation}
As it can be clearly seen, the advantage of this definition is that the momentum dependence of integrand is actually through the familiar form of propagator in free scalar filed theory.  So, analogous to loop computations in perturbation theory, we can analytically evaluate the momentum integrals here. The only difference is that due to using this method, besides performing every loop integral, a new integral over the unknown spectral density function appears as well.  In another word, the non perturbative aspect of each integral is encoded in an integral over the spectral density function, $\rh(\mu^2)$. In contrast to original integrals, the new integrals are not so problematic for numerical computations; it is simply because the spectral density function is an analytic function of just one variable which can be numerically integrated in Lorentz space. In the following sub we first develop using the method introduced above and leave Monte Carlo computation of the spectral density function as for the next subsections.

Let us return to equation (\ref{3.3}) and apply the spectral density function method to it. Relatedly, we have brought a careful computation of $Y(p^2)$ in (\ref{AppSunset}).
The final results reads as the following:
\begin{equation}\label{two.5}
\textbf{Y}(p^2)= \int_{0}^{\infty}\int_{0}^{\infty}\int_{0}^{\infty} d\mu_{1}^2 d\mu_{2}^2 d\mu_{3}^2\, \rh({\mu_{1}^2}) \rh({\mu_{2}^2})\rh({\mu_{3}^2})\, W(p^2,\mu_{1}^2,\mu_{2}^2,\mu_{3}^2)
\end{equation}
where the function $W(p^2,\mu_{1}^2,\mu_{2}^2,\mu_{3}^2)$ is defined as:
\begin{equation}
W(p^2,\mu_{i}^2)=i\frac{g^2}{6}\,\int_{0}^{1}\int_{0}^{1}\frac{x dx dy}{(4\pi)^4 \alpha^2} \left((\beta-\gamma p^2)\log\left(\frac{\beta-\gamma p^2}{\beta-\gamma m^2}\right)+\gamma(p^2-m^2)\right)
\end{equation}
with
\begin{equation}\label{two.7}
\begin{split}
&\alpha=\,(y-y^2-1)\,x^2+\,x,\\
& \beta=\,y \,x \,\mu_{1}^2+\,(1-y)\, x \,\mu_{2}^2+\,(1-x) \,\mu_{3}^2\\
& \gamma=\,x^2 \,(1-x)\,y\,(1-y)/\alpha.
\end{split}
\end{equation}
As it can be seen in (\ref{two.5}), besides a simply calculable function $W$, $\textbf{Y}(p^2)$ is constituted  of three spectral functions. It is expectable that substituting (\ref{two.5}) into
(\ref{3.1}) and simultaneously using the spectral representation method gives an integral equation for the spectral density function.  In fact the price that we must pay to get analytically performable loop integrals is to solve a numerical integral equation
for spectral function. We will refer to this equation as the "SD equation for the spectral density function". 
Deriving this equation from the Schwinger-Dyson equation of two-point function is the matter which we will discuss about  in detail in next subsection.

\subsection{Schwinger-Dyson Equation for Spectral Density Function }
We first remind that from (\ref{2.12}) the spectral density function can be expressed in term of full propagator as  $\rh(p^2)=\frac{1}{\pi}\, \textbf{Re}\, \textbf{G}(p^2)$. Therefore, by getting the real part of equation (\ref{3.1}), we obtain our desired integral equation as:
\begin{equation}\label{3.12}
\rh(p^2)=\delta^4(p^2-m^2)+\,\prod_{i=1}^{4}\,\int_{0}^{\infty} d\mu_{i}^2\, \rh({\mu_{i}^2})\,\,\,J(p^2,\mu_{i}^2;i\in \{1,2,3,4\})
\end{equation} 
where
\begin{equation}\label{3.99}
J(p^2,\mu_{4}^2,\mu_{i}^2;i\in \{1,2,3\})=\textbf{Re} \left(\frac{i G(p^2)\,\, W(p^2,\mu_{i}^2;i\in \{1,2,3\})}{\pi \,\,(p^2-\mu_{4}^2+i \epsilon)}  \right).
\end{equation}
Now, our main task is to numerically solve equation (\ref{3.12}). The equation is actually a non-linear integral equation which after being linearized, takes the familiar form of the scattering equation in quantum mechanics. Relatedly, the authors of \citep{bdmc1} have introduced an interesting Monte-Carlo method to solve such linear equations. As they have emphasized, by use of    their method, it would be possible to overcome some old problems in quantum field theory like the "singe problem".

In order to effectively use the method of \citep{bdmc1}, we start to solve equation (\ref{3.12}) through
"Newton's iteration". Newton's method in its own right deals with a linear recursive relation which in an integral fashion is similar to the integral equation studied in \citep{bdmc1}. So we first need to derive the Newton's recursive formula corresponded to (\ref{3.12}). To this end we rewrite (\ref{3.12}) as the following functional equation:
\begin{equation}
F[\rh(p^2)]:=\rho_{f}(p^2)+\,\prod_{i=1}^{4}\,\int_{0}^{\infty} d\mu_{i}^2 \rh({\mu_{i}^2})\,\,\,J(p^2,\mu_{i}^2;i\in \{1,2,3,4\})-\rh(p^2)=0;
\end{equation}
in this equation $\rho_{f}(p^2)=\delta^4(p^2-m^2)$ is the spectral density function of the 
free theory. Now we expand $F[\rh(p^2)]$ around a starting solution, i.e. $\rh_{n}(p^2)$ and find:
\begin{equation}\label{two.11}
\boxed{
\rh_{n+1}(p^2)=\rho_{f}(p^2)-3 \prod_{i=1}^{4}\int_{0}^{\infty}d\mu_{i}^2\,\rh_{n}(p^2)+
\int_{0}^{\infty}d\mu_{1}^2\,\rh_{n+1}(\mu_{1}^2)\,K(\mu_{1}^2,p^2)}
\end{equation}
with:
\begin{equation}
K(\mu,p^2)=\int_{0}^{\infty}\int_{0}^{\infty}\int_{0}^{\infty}d\mu_{2}^2d\mu_{3}^2d\mu_{4}^2\left(
J(p^2,\mu^2,\mu_{2}^2,\mu_{3}^2,\mu_{4}^2)+3\frac{}{}J(p^2\mu_{2}^2,\mu^2,\mu_{3}^2,\mu_{4}^2) \right).
\end{equation}
 We get $\rho_{f}(p^2)$ as the
starting solution and continue iterating. In next subsection we shall show the Monte Carlo solution of equation (\ref{two.11}) and compare it with the result of perturbation theory.

\subsection{Spectral Density Function, Monte Carlo Computation  Versus Perturbation Theory Result}
Before starting to discuss about the Monte Carlo solution of (\ref{3.12}),
let us give a few comments about the solution of perturbation field theory for this equation.
Firstly as it is usually taught in every course of quantum field theory,  two-point function and correspondingly spectral density function of the scalar theory are non-renormalizable up to first order in perturbative expansion. As a confirmation of this statement, we have seen that the function $J(p^2,\mu_{i}^2;i\in \{1,2,3,4\})$ in (\ref{3.12}) appears firstly from the second order of perturbation. The first non-vanishing loop contribution to spectral function, namely the perturbative two-loop contribution, can be simply
found by getting each of $\rh(p^2)$ functions in the integral term of (\ref{3.12}) equal to its tree level expression, namely $\delta^{(4)}(p^2-m^2)$.  As a result the spectral density function at second order turns out to be:
\begin{equation}\label{two.12+1}
\rh_{\text{per.}}(p^2)=\delta^4(p^2-m^2)-\,\textbf{Re}\left[i\frac{g^2}{6 \pi}\,\int_{0}^{1}\int_{0}^{1}\frac{x dx dy}{(4\pi)^4 \alpha^2} \left(\gamma(p^2-m^2)+(m^2-\gamma p^2)\log\frac{\frac{m^2}{p^2}-\gamma}{\frac{m^2}{p^2}(1-\gamma)}\right)\right]
\end{equation} 
According to (\ref{two.7}), $\gamma$ is a function of two  variables $x$ and $y$ which are both limited to change in $[0,1]$. Relatedly, the maximum value of this function is equal to $\frac{1}{9}$ at point $(x=\frac{2}{3},y=\frac{1}{2})$, so $\rh(p^2)$
has a branch cut beginning at  
$p^2=9 m^2$,
at the threshold for creation of three scalar particles. The point is that for $p^2<9 m^2$ the term $\textbf{Re}[\dots]$ in (\ref{two.12+1}) is always vanishing. As soon as $p^2$ exceeds the threshold,  the logarithm finds an imaginary part in a small region in the space of integral variables; consequently the term $\textbf{Re}[\dots]$ becomes non-vanishing through this region.\footnote{Graphically,  the following Feynman diagram shows a mediator particle may be so out of shell that can create three on-shell particles.

\begin{equation*}
\begin{fmffile}{Feynman/sunsetscat}
	\begin{fmfgraph*}(40,40)
		\fmfsurroundn{v}{12} 
		\fmfsurroundn{u}{6} 
	\fmfshift{(.6w,0)}{u1}
	\fmfshift{(1.1w,-.1h)}{u2}
	\fmfshift{(1.1w,.1h)}{u6}
	\fmfshift{(-.6w,0)}{u4}
	\fmfshift{(-1.1w,-.1h)}{u3}
	\fmfshift{(-1.1w,0.1h)}{u5}	
	\fmfshift{(-1.w,-0.27h)}{v6}	
	\fmfshift{(1.w,-0.27h)}{v2}	
	\fmf{dashes,tension=1}{v6,u4}
		\fmf{dashes,tension=1}{u3,u4}
		\fmf{dashes,tension=1}{u5,u4}
		\fmf{dashes,tension=1}{v2,u1}
		\fmf{dashes,tension=1}{u2,u1}
		\fmf{dashes,tension=1}{u6,u1}
	
	\fmf{fermion,label=$p$,l.s=right}{u4,v7}
	\fmf{fermion,label=$p$}{v1,u1}
	
	\fmf{plain,left=.97}{v7,v1}
	\fmf{plain,right=.97}{v7,v1}
	\fmf{plain}{v7,v1}

	
	\fmfshift{(.25w,.45h)}{v9}	
	\end{fmfgraph*}
\end{fmffile}
\end{equation*}
}

\begin{SCfigure}
  \centering
 \includegraphics[width=0.60\textwidth]
{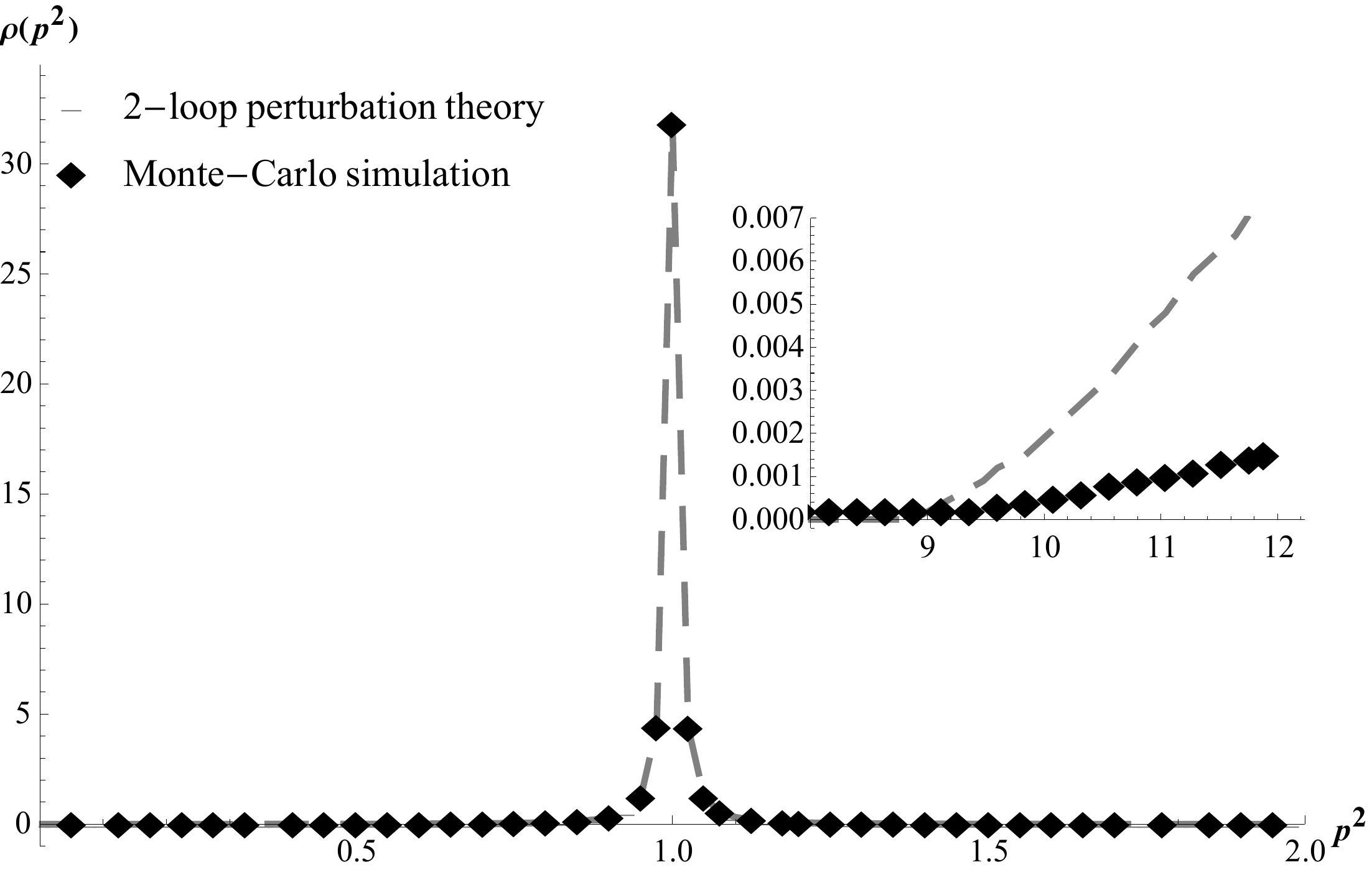}
  \caption{Comparison between spectral density function of $\phi^4$ theory obtained from Monte Carlo method with that of computed in perturbation theory. This plot has been depicted for $g=3(2\pi)^3$ (or equivalently for $\lambda_{eff}=3\pi/2>1$) and $m=1$. We have also chosen $\epsilon=0.01$ in both (\ref{3.99}) and the representation of delta function, namely in $\frac{\epsilon}{\epsilon^2+(p^2-m^2)}$.}\label{spectral}
\end{SCfigure}

The second comment is about the  the delta peak at $p^2=m^2$ shown in Fig.(\ref{spectral}). This  single peak simply demonstrates the only existing particle excitation of theory in the weak coupling regime. As it has been well-known, in full agreement with this observation,  the $\phi^4$-potential
is repulsive  in weak coupling regime and does not allow the particles to combine with each other and to make bound states. 
However, one might want to know how this result would be modified in non-perturbative
regime.



In Fig.(\ref{spectral}) we have also shown the result of Monte Carlo 
numeric. Although equation (\ref{3.12}) has been obtained via a particular truncation of SD equation, it is really a non-perturbative equation dominant not only in weak regime but also beyond that. Therefore the Monte Carlo plot in Fig.(\ref{spectral}) is actually a non-perturbative result. That to what extent of coupling constant this result remains reliable deserves more discussion which we leave it to the next section. However let us recall that a main goal in this paper is just  to develop the spectral function method for solving SD equations and not to find more precise results for a toy theory as $\phi^4$. 

In summary, as to answer the question asked in two previous paragraphs, we can say that the only difference between perturbative and non-perturbative behaviour of spectral function arises in the branch cut range.\footnote{In agreement with Monte Carlo result, it is straightforward but tedious to show that the function  $J(p^2,\mu_{i}^2)$ has in general a branch cut beginning at $p^2=9m^2$.  }  This result explicitly emphasises that $\phi^4$-potential will remain repulsive even beyond the weak regime, with the same threshold for creation of three particles as in perturbation theory.

\section{Computing the Scattering Amplitude Beyond the Weak Coupling Regime}\label{sec4}
In this section we are going to exploit the spectral density function obtained in previous section 
and compute the non-perturbative vertex function as the basis for computing the real-time scattering amplitude $\mathcal{M}(\phi \phi\rightarrow \phi \phi)$.

\subsection{Vertex Function in a Bold Expansion and the Pad\'{e} Approximation}
\label{VertexPade}
Similar to the case of inverse propagator $\Gam^{(2)}$, we now proceed to express the vertex function 
$\Gam^{(4)}$ in a bold expansion. However, for the sake of simplicity we restrict our following study to the lowest order of bold expansion, namely the second order in $\textbf{G}$ (see (\ref{1.8})).

 Imposing the renormalization
condition (\ref{rc33}),
one can simply find $Z_{g}$ and consequently the renormalized vertex function up to above-mentioned order. The result may be given as:
\begin{equation}\label{three.1}
\begin{split}
&\Gam^{(4)}(p_{1},p_{2},p_{3},p_{4})=\,ig\\
&-\,\frac{ig}{2}\left(\int \frac{d^4k\,d^4q}{(2\pi)^8}\textbf{G}(k^2) \textbf{G}(q^2)\,(\Gam^{(4)}(p_{3},p_{4},q,k)-\tilde{\Gam}^{(4)}(p_{3},p_{4},q,k)) +\,(2 \leftrightarrow 3,4)\right)
\end{split}
\end{equation}
 where $k^{\mu}=p_{1}^{\mu}+p_{2}^{\mu}-q^{\mu}$ and $\tilde{\Gam}^{(4)}$ denotes that the vertex function  being evaluated at the renormalization point (\ref{1.20}).
At first look it might be thought that the equation (\ref{three.1}) is an integral equation analogous 
to equation (\ref{3.1}). 
Although those are either non-linear integral equations but there is  a delicate point which clearly distinguishes them from each other. The point is that in contrast to (\ref{3.1}), even by using spectral method we would not be able to get rid of momentum integrals in (\ref{three.1}). It is simply because there has not been known any integral representation for vertex function in term of its tree level expression. Alternatively, one may think to find a full numerical solution for (\ref{three.1}), something which is principally possible but hard to implement. The reason for that is the vertex function, as was previously pointed out, is a function of six Lorentz scalars; so discretizing the associated six dimensional phase space produces so large number of discrete points that related Monte Carlo implementations would not be possible through utilizing common computational resources.  

To remedy the problem, we resort a well-known approximation method in mathematics, namely \textbf{Pad\'{e} approximation}  of a function. Pad\'{e} approximant is the approximation of a function by a rational function of given order. Given the degree of polynomials in numerator and denominator, $m$ and $n$, and the point $a$ around which the approximation is requested, associated Pad\'{e} approximant $[m/n,a]_{f}(x)$ will be determined just by demanding the approximant's power series be the same as original function's power series \citep{pade}.  In Appendix(\ref{AppPade}) we have written a number of first low-order Pad\'{e} approximants for an arbitrary function $f(x)$ around $x=0$.\footnote{For simplicity, we show the Pad\'{e} approximant around $x=0$ as $[m/n]_{f}(x)$. } As an important point, It is well-known that Pad\'{e} approximant of a function gives better approximation than its truncated Taylor series. By this motivation we will try to exploit this technique to solve linear integral equations. Before applying this technique to our specific integral equation (\ref{three.1}), let us consider a general linear integral equation as:
$f(x)=g(x)+\int_{a}^{b}f(y)K(x,y)dy$. 
The idea of using Pad\'{e} approximation to solve this equation is to insert a new variable $\epsilon$ multiplying the integral term in RHS and consequently promote the function $f(x)$ to be a function of both $x$   and $\epsilon$, namely $f(x,\epsilon)$. This leads the integral equation takes the following form:
\begin{equation}\label{three.2}
f(x,\epsilon)=g(x)+\,\epsilon\int_{a}^{b}f(y,\epsilon)K(x,y)dy.
\end{equation}
Considering $f(x,0)=g(x)$, in order to obtain Pad\'{e} approximant of $f(x,\epsilon)$ around $\epsilon=0$ one just needs to have $\partial_{\epsilon}^{(n)}f(x,\epsilon)\vert_{\epsilon=0}$.                                         Interestingly,   since these derivatives are evaluated at $\epsilon=0$, $f(y,\epsilon)$ in the integral will always be equal to $g(x)$ and consequently the integral will be simply performed. As the final step we must compute
the Pad\'{e} approximant  at the point $\epsilon=1$ which gives nothing but Pad\'{e} approximate solution for $f(x)$. 
Using this technique, in (\ref{AppComp}) we have considered a toy equation with an exact solution and have solved it approximately. There has been shown that higher order Pad\'{e} approximate solutions tend to quickly converge to the exact solution.   
  
Following the procedure developed in previous paragraph, let us now return to equation (\ref{three.1}) and apply the $[0/1]$ Pad\'{e} approximant to it. Considering the formula given in (\ref{AppPade}), the only necessary objects for computing $\Gam^{(4)}_{[0/1]}$ are $\Gam^{(4)}(p_{1},p_{2},p_{3},p_{4},\epsilon=0)$ and $\partial_{\epsilon}\Gam^{(4)}(p_{1},p_{2},p_{3},p_{4},\epsilon)\vert_{\epsilon=0}$. While he former is simply equal to $ig$, the latter is given by:
\begin{equation}\label{three.3}
\begin{split}
&\partial_{\epsilon}\Gam^{(4)}(p_{1},p_{2},p_{3},p_{4},\epsilon)\vert_{\epsilon=0}=\,\,\,ig\,X(p_{1},p_{2},p_{3},p_{4})=\\
&-\frac{i g^2}{32 \pi^2} \prod_{i=1}^{2}\int_{0}^{\infty}
 d\mu_{i}^2 \rh(\mu_{i}^2)\,\int_{0}^{1}dx\log\left(\frac{(x\mu_{1}^2+(1-x)\mu_{2}^2-4m^2(x-x^2))(x\mu_{1}^2+(1-x)\mu_{2}^2)^2}{\prod_{m=2}^{4}(x\mu_{1}^2+(1-x)\mu_{2}^2-(p_{1}+p_{m})^2(x-x^2))}\right)\end{split}
\end{equation}
where $X(p_{1},p_{2},p_{3},p_{4})$ can be simply read by dropping an $ig$ factor from the  expression in second line.  As a result,
the $[0/1]$ Pad\'{e} approximate solution of equation (\ref{three.1}) is as the following:
\begin{equation}\label{three.4}
\Gam^{(4)}_{[0/1]}(p_{1},p_{2},p_{3},p_{4})=\,\frac{i g}{1-X(p_{1},p_{2},p_{3},p_{4})}.
\end{equation} 
Note that to find the vertex function (\ref{three.4}) we have to utilize  our previously found non-perturbative spectral density function. Using the result given in Fig.(\ref{spectral}),   in Fig.(\ref{ReIm}) we have plotted the real and imaginary parts of $\Gam^{(4)}_{[0/1]}(s,t=0,u=0)$ and have also compared each of them with the perturbative result at the same point. By the one-loop perturbative vertex function we mean  $\Gam^{(4)}_{\text{per.}}=ig(1+\hat{X})$ where $\hat{X}$ is the same as $X$ but evaluated for the one-loop perturbative spectral function, namely $\rh(p^2)=\delta^{(4)}(p^2-m^2)$. 

Let us emphasis that although (\ref{three.4}) is basically a  non-perturbative solution for truncated SD equation (\ref{three.1}), for two reasons $\Gam^{(4)}_{[0/1]}$ and $\Gam^{(4)}_{\text{per.}}$ are nearly the same: firstly, $\rh_{\text{per.}}(p^2)$ and $\rh_{\text{non-per.}}(p^2)$  are the same in a wide range of momentum as depicted in Fig.(\ref{spectral}). The other reason is, (\ref{three.4}) may be regarded as the one-loop resummed expression of vertex function for sufficiently small values of $g$. However, In the last part of next subsection we discuss the  unitarity bound may allow $\Gam^{(4)}_{[0/1]}$ to be valid for a wider range of $g$ in comparison with $\Gam^{(4)}_{\text{per.}}$.   It would mean that we have found a non-perturbative solution applicable beyond the
weak coupling regime.
Considering this fact that  (\ref{three.4}) is individually an approximate solution, one may ask how much this solution may be  really reliable. We are going to answer to this question in next subsection.

%
%
\subsection{Higher Order Pad\'{e} Approximants of Vertex Function}
Practically, in order to find a Pad\'{e} approximate solution we must continue finding the higher order  Pad\'{e} approximants as long as a sign of convergence appears in the successive approximants.  
In a general integral equation like (\ref{three.2}), it is simply possible to keep track of when the successive 
approximants are converging. Compared to the leading order, the only difference which arises in higher orders is appearance of some simply performable multi-integral terms like $\int_{a}^{b}\int_{a}^{b}g(z)K(z,y)K(y,x)dz dy$ through computations. In our specific equation, i.e. (\ref{three.1}), however, renormalization is a vital point which seriously affects the higher order approximants. In order to clarify it, let us rewrite the second partial derivative of
$\Gam^{(4)}(p_{1},p_{2},p_{3},p_{4},\epsilon)$ with respect to $\epsilon$, as the basis for computation of $\Gam^{(4)}_{[0/2]}$ or $\Gam^{(4)}_{[1/1]}$, as the following :
\begin{equation}\label{three.5}
\partial^2_{\epsilon}\Gam^{(4)}(p_{1},p_{2},p_{3},p_{4},\epsilon)\vert_{\epsilon=0}=
-\,ig\int \frac{d^4k\,d^4q}{(2\pi)^8}\underbrace{\textbf{G}(k^2) \textbf{G}(q^2)\,X(p_{3},p_{4},q,k)}_{k+q=p_{1}+p_{2}}+\,(2 \leftrightarrow 3,4).
\end{equation}
According to (\ref{three.3}), $X(p_{3},p_{4},q,k)$ is a finite-valued function; this means that the integral in (\ref{three.5}) is definitely divergent because of lacking a similar integral contribution which can subtract the divergent part of dressed propagators. This divergence has a direct outcome:
our Pad\'{e} approximation is not consistent with renormalization process. Consequently, in order to obtain higher order Pad\'{e} approximants, we must go beyond the leading order in $\textbf{G}$ in (\ref{1.7}) and consider the terms including more number of $\Gam^{(4)}$'s.  
To this end we may write  (\ref{1.7}) as
\begin{equation}\label{three.6}
\Gam^{(4)}(p_{1},\dots,\epsilon)=i\,Z_{g}\,g\,(1+\,\#\,\epsilon\int \textbf{G}\textbf{G}\Gam^{(4)}+\,\#\,\epsilon^2\int \textbf{G}\textbf{G}\textbf{G}\textbf{G}\Gam^{(4)}\Gam^{(4)}+\dots)
\end{equation}
where the power of $\epsilon$ in each term really counts the number of full vertices.  One may naturally expect to find new contributions to $\partial^2_{\epsilon}\Gam^{(4)}$ when taking into account the second term of (\ref{three.6}) to regularize (\ref{three.5}). This is what will really happen if one determines $Z_{g}$ by using associated renormalization condition. In this paper however, we are not interested in doing such lengthy computations for a toy theory as $\phi^{4}$.  We just want to emphasis that this is essentially possible to compute higher order derivatives of $\Gam^{(4)}$ and to find higher order Pad\'{e} approximants respectively.

\begin{figure}
  \centering
  \includegraphics[scale=.30]{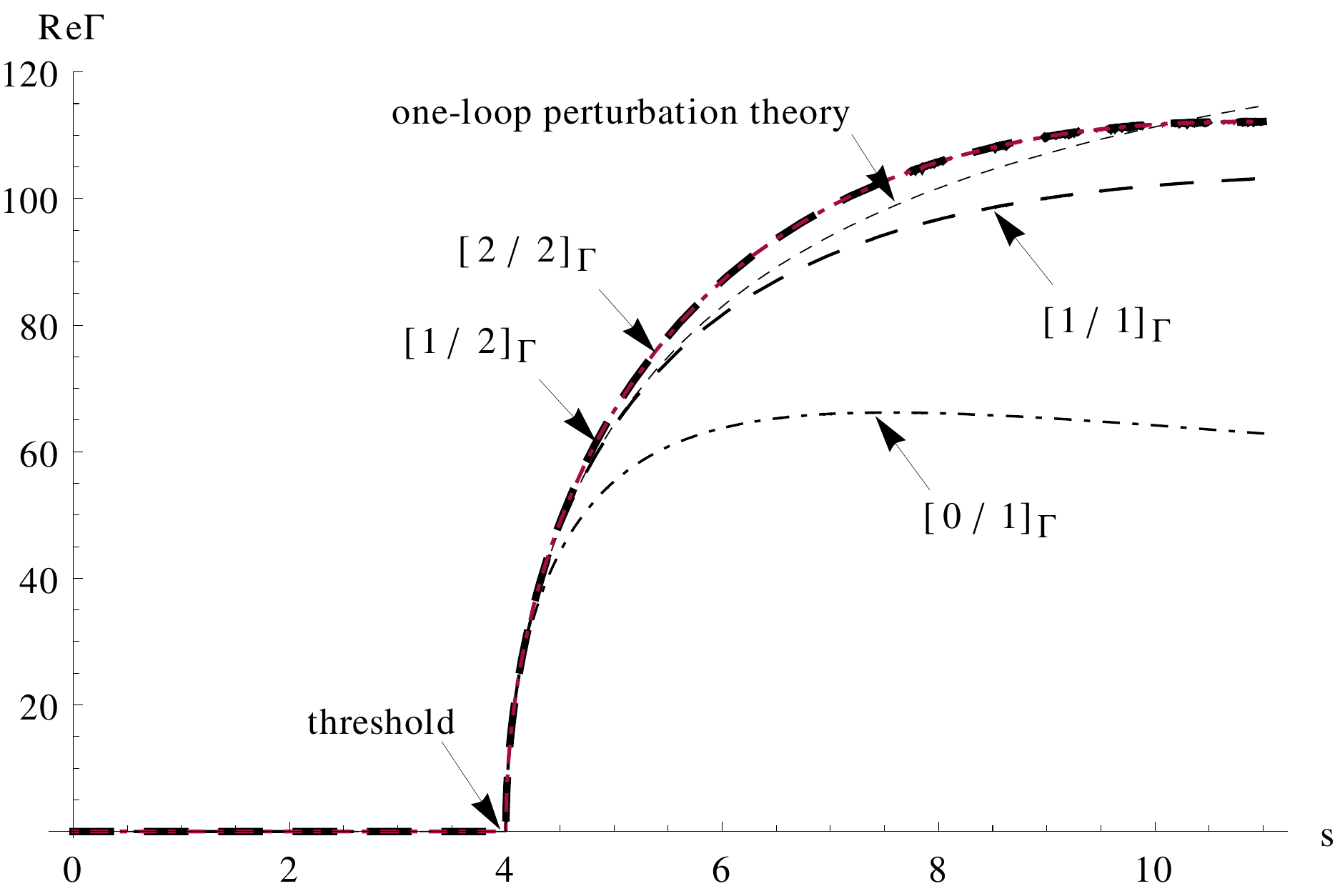}
  \hspace{2cm}
  \includegraphics[scale=.30]{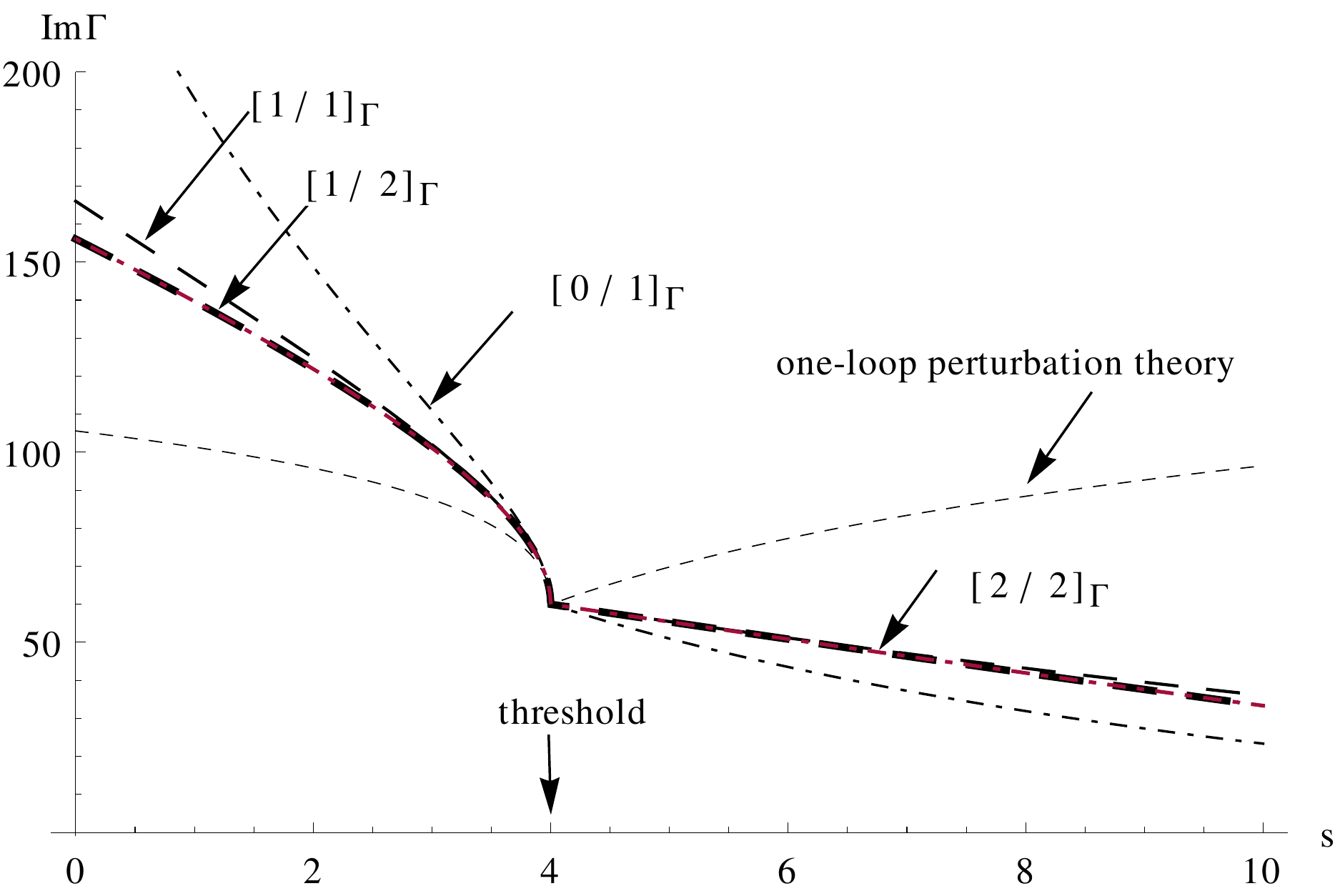}
  \caption{In left panel we have depicted the real part of $\Gam$ for $g=85$ (or equivalently $\lambda_{eff}=0.53$) while the imaginary part shown in right panel is related to $g=60$ (or $\lambda_{eff}=0.38$).
  Unitarity bound constrains the one-loop perturbation results to be valid just for $g\lesssim33$ \cite{Luscher:1987ay}. Therefore, the one-loop curves in both panels are invalid. (We have also scaled energies by choosing $m=1$.)} 
  \label{ReIm}
  \end{figure}

Instead of rigorous computations, we now try to make it clear qualitatively that how higher order Pad\'{e} approximants may converge to a non-perturbative solution for (\ref{three.6}).
  As an order estimation, using $\int \textbf{G}\textbf{G}\,ig=\,-2 X$ (see (\ref{three.3})), $\partial^2_{\epsilon}\Gam^{(4)}$ turns out to be as the following (see (\ref{three.5})):
 \begin{equation}
 \partial^2_{\epsilon}\Gam^{(4)}(p_{1},p_{2},p_{3},p_{4},\epsilon)\vert_{\epsilon=0}\sim\,2ig X^2(p_{1},p_{2},p_{3},p_{4})
 \end{equation} 
and similarly:
 \begin{equation}
 \partial^3_{\epsilon}\Gam^{(4)}(p_{1},p_{2},p_{3},p_{4},\epsilon)\vert_{\epsilon=0}\sim\,6ig X^3(p_{1},p_{2},p_{3},p_{4}).
 \end{equation} 
The estimation used here is based on counting both the number of bold propagators and vertices.\footnote{In order to find the accurate value of above derivatives, we must perform all two- and three-loop integrals.}
Considering expressions above as the input for the Pad\'{e} approximants of vertex function,
we have utilized the formula given in (\ref{AppPade}) and have found a number of approximate solutions of $\Gam^{(4)}$ as well. In Fig.(\ref{ReIm}), we have plotted the real and imaginary parts of resultant approximants for the $s$ channel scattering when $t$ and $u$ channels have been turned off. As it can be clearly seen in this figure, the successive approximants converge so quickly that
$[2/2]_{\Gam}$ may be regarded as a non-perturbative solution of the vertex function. However, that to what extent of coupling constant $[2/2]_{\Gam}$ would be a reliable solution is a different matter which should be discussed through another arguments like "\textbf{unitarity bound}". As it is well-known, one of the ways through which one can constrain the quantum amplitudes is imposing unitarity bound to the vertex function. The associated result would be a range for the coupling constant wherein the computation remains valid as well. However we do not  want to apply this bound to the higher order Pad\'{e} approximate solutions because, our convergent solution of vertex function is an estimate one itself. We just reasonably claim that if one compute
two- and three-loop integrals in perturbation theory, she will be able to find convergent Pad\'{e} solutions and to constrain them via untarity bound.  In this regard, we leave more precise computations to the case of QCD in our next work.

Before ending this subsection it may be necessary to note that our developed Pad\'{e} approximation in (\ref{three.6}) is actually like a perturbation on the vertex function. In another word, the new variable $\epsilon$ appears as a multiplying factor in front of every $\Gam^{(4)}$ and its power is equal to the number of multiplying $\Gam^{(4)}$'s. It is why our Pad\'{e} solutions distinguishes from the perturbation theory result.

\section{Conclusion}\label{sec5}
Our results reported in this paper may be encountered with several follow-up questions. We list 
some of them below and leave the answers to our future work.

Firstly let us recall that in this paper we have constructed a new method to extract non-perturbative information from a field theory beyond its weak coupling regime. Our method is based on numerically solving the Schwinger-Dyson equation of spectral density function through Bold Diagrammatic Monte Carlo method.
The desired SD equation was obtained by combining the SD equations of vertex and two-point function in a particular way.
In more details, we first substituted the truncated vertex equation into the SD equation of two-point function to derive a non-linear integral equation for the two-point function.
We then showed that by making use of the spectral representation, the momentum integral in this equation could be performed analytically, resulting in an integral equation for the spectral density function. 

As a natural extension, one may be interested in improving the results given in this paper.
The first suggestion to improve our results would be undoubtedly to hold more number of terms in truncated vertex equation. However, this leads to a more complicated integral equation for spectral function and as a result, one needs to implement more advanced numerical computations.    

In another direction one can improve the results by the idea of applying this method to the gauge field theories case. Every typical gauge theory, like QCD, is equipped with a continuous gauge group, like $SU(N)$. 
As it is well known, the "\textbf{Large-$N$ expansion}" is a particular perturbative technique in such theories.
Inspired by this fact, one may consider an $N$-component scalar field enjoying the $O(N)$ internal symmetry.
Obviously, in two well-defined limits, the $O(N)$ scalar theory  is perturbatively soluble; firstly in the IR regime around the zero coupling point and secondly in the large-$N$ limit mentioned above. Although some types of methods are not capable to 
capture contributions beyond the leading order in $1/N$ expansion, the large-$N$ Schwinger-Dyson equations may self-consistently produce the sub-leading contributions as well \citep{zinn}.
Similarly, in the current paper, we have exploited the power of Schwinger-Dyson equations 
to find non-perturbative solutions beyond the IR regime too. As a result, one might be tempted to combine the results of  large-$N$ expansion  with those of BDMC (firstly introduced in present paper) to find an interpolating solution for the $O(N)$ scalar theory. 

  \begin{figure}
  \centering
  \includegraphics[scale=.39]{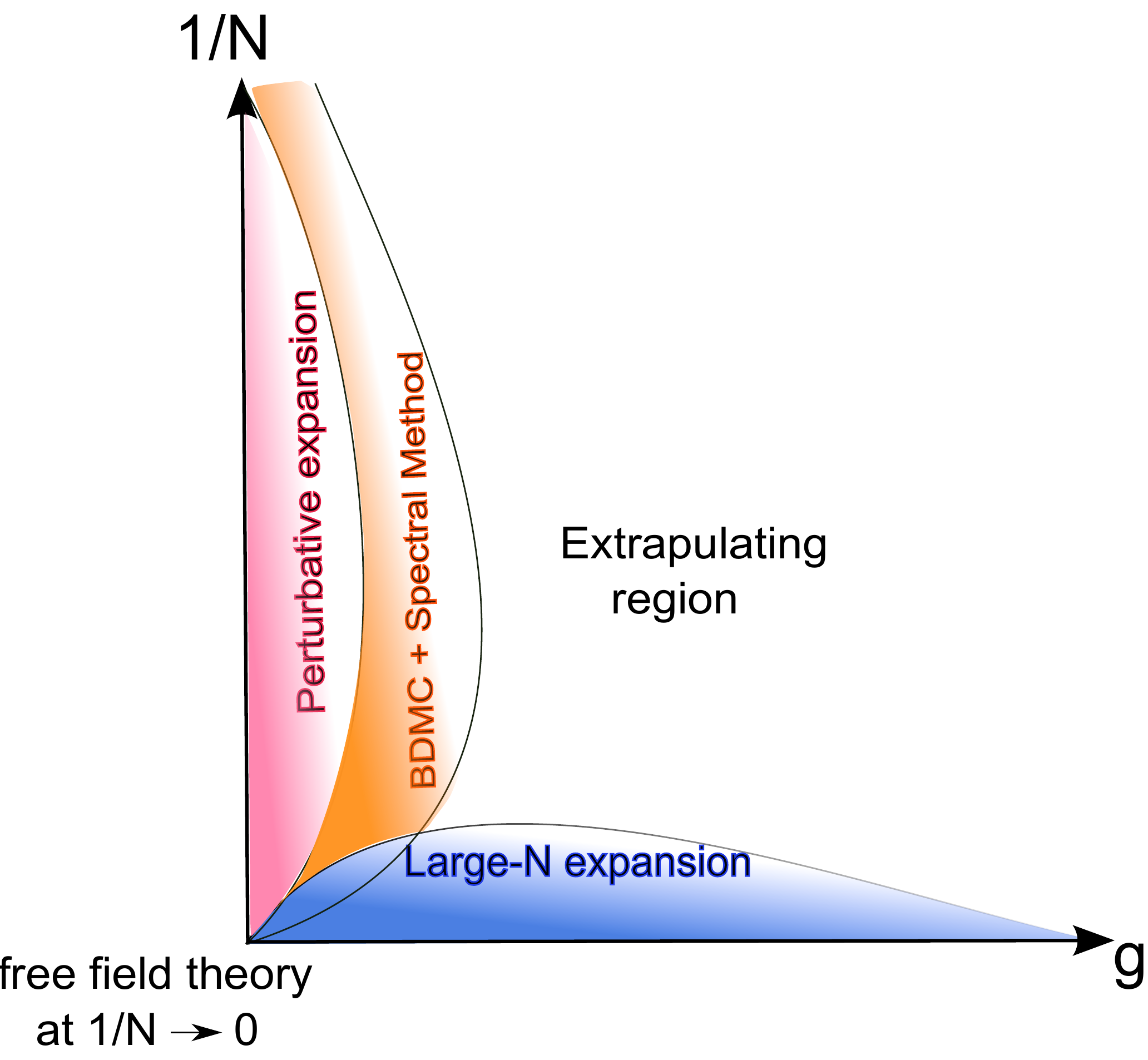}
  \caption{Indication of regimes of the validity of the method developed in this paper and the large-$N$ expansion
  in the two dimensional parameter space of $O(N)$ scalar filed theories, labeled by the renormalized coupling constant $g$ and the inverse number of colors $1/N$.} 
  \label{ParameterSpace}
  \end{figure} 

To be more clarified we have illustrated both perturbative regimes discussed above in a two dimensional parameter space in Fig.(\ref{ParameterSpace}). As it has been shown, BDMC combined with the spectral method has made us capable of covering a more extended region in the parameter space. When interpolating the results, one would be actually able to explore the theory for both larger values of coupling constant and finite values of $N$; something which seems really salient in a gauge filed theory like QCD.  

Using the combination of BDMC with our spectral method when studying QCD has another valuable advantage too. Since we use dimensional regularization to regularize momentum integrals, gauge symmetry will not be broken and consequently our numeric will not suffer from the old problems related to presence of a cut-off in integrals. It can also increase the efficiency of Monte Carlo numeric. 

Relatedly, it would be interesting to apply the method developed in this paper to the case of $\mathcal{N}=4$ SYM filed theory with $SU(N_{c})$ gauge group. As it is well known, AdS/CFT duality is able to describe planar, strongly coupled limit of the SYM theory via using classical supergravity. During recent years, it has been made many efforts to compare in strong coupling regime of the SYM theory with the low energy QCD \cite{Gubser:2006qh}. As a reasonable improvement, by finding an interpolating solution between AdS/CFT results and those that can be obtained through applying the method of this paper to SYM theory, one can reach to more precise predictions for strong QCD as a gauge theory with finite $N_{c}$. 

In a different direction from the paths sketched above, our method makes it possible to nonperturbatively study of a relativistic field theory with a complex action, like QCD at finite chemical potential. Using non-perturbative lattice simulations, it has been shown that the complex scalar theory, namely self-interacting $(\phi \phi^*)^2$ theory, in $d=4$ represents a transition to a condensed phase at large baryon density \cite{Aarts:2008wh}. 
It would be interesting to investigate whether BDMC can reproduce the same phase transition when applying to
SD equation of spectral density function in complex scalar theory.

Besides probable generalizations mentioned above, it would be also possible to use the method of current paper in the context of classical physics. A very interesting place to explore is homogeneous, isotropic turbulence, HIT. Turbulence is the chaotic motion of a fluid when shearing by an external force. However, the statistical theory of turbulence,   firstly begun by Reynolds, suffers from the familiar statistical closure problem. This problem may be briefly reviewed as it follows. Consider the one dimensional Navier-Stokes equation as:
\begin{equation}
  \mathcal{L}_{0}\,U:=\,(\partial _{t}-\nu_{0}\,\partial^2_{ x})\,U=\,M\,U\,U   
\end{equation}
with $\mathcal{L}_{0}$ the linear operator and $M$ the operator including the effect of pressure
and non-linear term. According to  Reynolds decomposition, it is necessary to decompose the velocity filed to a mean velocity $\langle U\rangle$ and a randomly fluctuating filed about the mean $u$, i.e. $U=\langle U\rangle+u$.
However when substituting the Reynolds decomposition into the Navier-Stokes equation we encounter with a hierarchy of moment equations as the following:
\begin{equation}
\begin{split}
  \mathcal{L}_{0}\,\langle U  \rangle&=\,M\,\langle uu \rangle\\
    \mathcal{L}_{0}\,\langle uu \rangle&=\,M\,\langle uuu \rangle\\
      \mathcal{L}_{0}\,\langle uuu \rangle&=\,M\,\langle uuuu \rangle\\
              ....&.........
\end{split}
\end{equation}  
Obviously we always have $N$ equations for $N+1$ unknown variables $\langle uu \rangle$, $\langle uuu \rangle$,$\langle uuuu \rangle$ , $\dots$\,. This is the closure problem of turbulence.

The problem mentioned above may be reformulated in the language of quantum field theory as well.
Before proceeding further let us rewrite the Navier-stokes equation in terms of the fluctuating field in the wave number space as :
\begin{equation}
\mathcal{L}_{0k}\, u_{k}=\,M_{k}\,u_{j}u_{k-j} 
\end{equation}
with the time variable implicitly considered.
As it is well known if one makes all the variables in Navier-Stokes equation dimensionless, the relevant Reynolds 
number will appear as a factor in front of the non-linear term.
So the Navier-Stokes equation with a stirring force term $f_{k}$ may be rewritten as:
\begin{equation}\label{app4.5}
\mathcal{L}_{0k}\, u_{k}=f_{k}+\,\lambda\,M_{k}\,u_{j}u_{k-j} .
\end{equation}  
Here, $\lambda$ is a book-keeping parameter which can be replaced by a Reynolds number after an appropriate scaling of variables. The really interesting point  is that when the book-keeping parameter increases from $\lambda=0$ to  
$\lambda=1$, the linear system changes to a non-linear one as well. The resultant non-linear system is actually  a turbulent flow. 
The solution of (\ref{app4.5}) may be formally given as a perturbative expansion on $\lambda$:
\begin{equation}
u_{k}=u_{k}^{(0)}+\lambda\,u_{k}^{(1)}+\lambda^2\,u_{k}^{(2)}+...
\end{equation}
where all $u^{(n)}_{k}$ terms are written in terms of $M_{k}$, $u_{k}^{(0)}$ and 
the "\textbf{Response function}" $R^{(0)}_{k}=\mathcal{L}_{0k}^{-1}$. As a result, the exact "\textbf{Covariance}" of the fluctuating velocity $C_{k}=\langle u_{k}u_{-k} \rangle$ and also the
exact response function $R_{k}$ may be perturbatively given by a pair of coupled, non-linear, integral equations.
In 1959 "Kraichnan" proposed that one could simply solve the closure problem just by truncating these coupled equations and simultaneously replacing $C^{(0)}_{k}$ and $R^{(0)}_{k}$ with $C_{k}$ and $R_{k}$ respectively.  
It is straightforward to show that the close set of "renormalized" Kraichnan equations is given by \citep{McComb}
\begin{equation}
\begin{split}
\mathcal{L}_{0k}C_{k}&=F_{k}+\,L(k,j)[R_{k}C_{j}C_{l}-R_{j}C_{l}C_{k}]\\
\mathcal{L}_{0k}R_{k}&=\delta(t-t')+L(k,j)R_{l}C_{j}.
\end{split}
\end{equation} 
Interestingly, these equations are analogous to the Schwinger-Dyson equations in quantum filed theory.
Similar to the Feynman diagrammatic representation of SD equations, the Kraichnan equations may be represented graphically
through the Wyld diagrams. By this motivation, it would be of most interest to use the method introduced
 in current paper to solve the coupled  Kraichnan equations. It would be really  interesting to explore that whether the non-perturbative solution of covariance can reproduce the famous "Kolmogorov" results on HIT. If so, we will have achieved a new analytic tool to investigate HIT. It will then be important to investigate probable new universal relations in HIT. We leave more study on this issue to our future work.

\acknowledgments
We would like to thank Amin Akhavan for useful discussion.

\appendix
\section{Appendices}
\label{Appendix}

\subsection{Derivative Terms in Schwinger-Dyson Equation of Vertex}
\label{AppDer}
In the following we derive a typical derivative term including both the lowest number of $\Gam^{(4)}$ and $\textbf{G}$ in the expansion of (\ref{1.7}). Clearly such term is produced by $\mathcal{D}$ in which $\Gam^{(4)}$ again, has to be also gotten with both the lowest number of $\Gam^{(4)}$ and $\textbf{G}$. Consistent with these considerations,   the only constituents of $\Gam^{(4)}$
which may contribute to functional derivative term would be $\mathcal{A}$ (see  (\ref{1.8}) and (\ref{1.15})). In the original notation of \citep{Pelster:2003rc}, we can formally rewrite $\mathcal{A}$ as ($i,j\in \{1,2,3,4\}$):\footnote{Note that $\textbf{G}(p,q)=(2 \pi)^4 \delta^4(p+q)\,\textbf{G}(p^2)$ and $V(p,q,k,l)=i \, Z_{g}g\,(2 \pi)^4 \delta^4(p+q+k+l)$.}
\begin{equation}
\begin{split}
\mathcal{A}(p_{i}.p_{j})=-\frac{1}{2}\int_{p_{5},p_{6},p_{7},p_{8}}V(p_{1},p_{2},-p_{5},-p_{6})\textbf{G}(p_{5},p_{7})\textbf{G}(p_{6},p_{8})&\Gam^{(4)}(-p_{7},-p_{8},p_{3},p_{4})\\
&+(2\leftrightarrow 3)+(2\leftrightarrow 4)
\end{split}
\end{equation} 
Differentiating the above expression with respect to $\mathcal{A}$ and then integrating the delta function, we obtain a one-loop expression. In the compact notation of (\ref{1.8}) it may be written as:
\begin{equation}\label{fashion}
\frac{\delta\mathcal{A}(p_{5},p_{2},p_{3},p_{4})}{\delta \textbf{G}(-p_{6},-p_{7})}=\,-\,\underbrace{(iZ_{g} g) \,\,\,\,\,\,\,\textbf{G}(p_{8}^2)\,\,\,\,\,\,\,\,\Gamma^{(4)}(p_{7},p_{8},p_{3},p_{4})}_{p_{6}^{\mu}=-p_{5}^{\mu}-p_{2}^{\mu}-p_{8}^{\mu},\,\,\,\,\,p_{7}^{\mu}=-p_{3}^{\mu}-p_{4}^{\mu}+p_{8}^{\mu}}\,\,\,\,+\,\,(2\leftrightarrow 3,4)
\end{equation}
Now we use (\ref{1.15}) to compute the desired functional derivative term:
\begin{equation}
\begin{split}
\frac{\delta{\Gam^{(4)}}(p_{5},p_{2},p_{3},p_{4})}{ \delta\textbf{G}(-p_{6},-p_{7})}&=
(i\,Z_{g} g)\,\frac{\delta\mathcal{A}(p_{5},p_{2},p_{3},p_{4})}{\delta \textbf{G}(-p_{6},-p_{7})}\\
&=
-\,\underbrace{(iZ_{g} g)^2 \,\,\,\,\,\,\,\textbf{G}(p_{8}^2)\,\,\,\,\,\,\,\,\Gamma^{(4)}(p_{7},p_{8},p_{3},p_{4})}_{p_{6}^{\mu}=-p_{5}^{\mu}-p_{2}^{\mu}-p_{8}^{\mu},\,\,\,\,\,p_{7}^{\mu}=-p_{3}^{\mu}-p_{4}^{\mu}+p_{8}^{\mu}}\,\,\,\,+\,\,(2\leftrightarrow 3,4)\end{split}
\end{equation}
Similarly one can continue the process and derive higher order terms made out of functional derivative terms.

\subsection{Computing the Sunset Diagram}
\label{AppSunset}
Equation (\ref{3.3}) may be rewritten as:
\begin{equation}\label{3.4}
\begin{split}
\Sig(p^2)&=\frac{i g^2}{6}\,\,\,\,\,\,\,\,\,\,\,\,\,\,\,\,\parbox{3mm}{
\begin{fmffile}{Feynman/sunsetscat2}
	\begin{fmfgraph*}(35,39)
		\fmfsurroundn{v}{12}  
		\fmfsurroundn{u}{6} 
	\fmfshift{(.6w,0)}{u1}
	\fmfshift{(1.1w,-.1h)}{u2}
	\fmfshift{(1.1w,.1h)}{u6}
	\fmfshift{(-.6w,0)}{u4}
	\fmfshift{(-1.1w,-.1h)}{u3}
	\fmfshift{(-1.1w,0.1h)}{u5}	
	\fmfshift{(-1.w,-0.27h)}{v6}	
	\fmfshift{(1.w,-0.27h)}{v2}	
%
	\fmf{fermion,label=$p$,l.s=right}{u4,v7}
	\fmf{fermion,label=$p$}{v1,u1}
	\fmf{fermion,left=.97,label=$k$}{v7,v1}
	\fmf{fermion,right=.97,label=$l$,l.s=left}{v7,v1}
	\fmf{fermion,label=$q$}{v7,v1}
	\fmfshift{(.25w,.45h)}{v9}	
	\end{fmfgraph*}
\end{fmffile}}
\\
&\,\,\,\,\,\,\,\,\,\,\,\,\,\,\,\,\,\,\,\,\,\,\,\,=\frac{i g^2}{6} i^3 \prod_{i=1}^{3}\int_{0}^{\infty} d\mu_{i}^2 \rh({\mu_{i}^2})\int\frac{d^4 k d^4 q}{(2 \pi)^8}\frac{1}{k^2-\mu_{1}^2}\frac{1}{q^2-\mu_{2}^2}\frac{1}{l^2-\mu_{3}^2}
\end{split}
\end{equation}
where $ l^{\nu}=-p^{\nu}-k^{\nu}-q^{\nu}$.
The RHS of this equation is a two-loop momentum integral containing three fractions in the integrand multiplied by each other. This integral which represents the sunset diagram is basically divergent. In what follows we firstly squeeze the denominator factors into a single expression and then use the dimensional regularization to extract the infinity. 
Using the following formula:
\begin{equation}
\prod_{i=1}^{3}\frac{1}{(k^2-\mu_{1}^2+i \epsilon)}\frac{1}{(q^2-\mu_{2}^2+i \epsilon)}\frac{1}{(l^2-\mu_{3}^2+i \epsilon)}=\int_{0}^{1}\int_{0}^{1}\int_{0}^{1} dx dy dz \delta(x+y+z-1)\frac{2}{D^3},
\end{equation}
 where $D=xk^2+ y q^2 + z (p+k+q)^2-x_{i}\mu_{i}^2+i \epsilon$ and $x_{i}\mu_{i}^2=x \mu_{1}^2+y \mu_{2}^2+z \mu_{3}^2$, one can write down the momentum integrals as:
 \begin{equation}\label{5.2}
 \begin{split}
 \int \int \frac{d^4k\,d^4q}{(2 \pi)^8}\,\frac{2}{D^3}&=\,\int \int \frac{d^4K\,d^4Q}{(2 \pi)^8}\,\frac{2}{(\alpha K^2+\beta Q^2+\gamma p^2-x_{i}\mu_{i}^2+i \epsilon)^3}\\
 &=\,\int \int \frac{d^4K_{E}\,d^4Q_{E}}{(2 \pi)^8}\,\frac{2}{(\alpha K_{E}^2+\beta Q_{E}^2+\gamma p^2+x_{i}\mu_{i}^2-i \epsilon)^3}\\
&= \,\lim_{n\rightarrow 4}\mu^{2(4-n)}\int \int \frac{d^nK_{E}\,d^nQ_{E}}{(2 \pi)^{2n}}\,\frac{2}{(\alpha K_{E}^2+\beta Q_{E}^2+\gamma p^2+x_{i}\mu_{i}^2-i \epsilon)^3}.
 \end{split}
\end{equation}  
 In first line we have linearly transformed $(k,q)$ to $(K,Q)$.
 Then in the second line, we have applied a Wick rotation ($d^4K\rightarrow i d^4K_{E}$) and finally in third line, we have started to dimensionally regularize the integral. Ley us also denote  that in the expressions given above $\mu$ is a quantity with the mass dimension which will be
disappeared after renormalization. We also have:
\begin{equation}
\alpha=x+z,\,\,\,\,\,\,\,\,\,\,\,\beta=\frac{xy+yz+zx}{\alpha},\,\,\,\,\,\,\,\,\,\gamma=\frac{xyz}{\alpha}.
\end{equation}
 Using $2/D^3=\int_{0}^{\infty}dt\, t^2\, e^{-Dt}$, one can separate
the integrals over $K$ and over $Q$ as well. After evaluating two Gaussian Integrals we obtain:
\begin{equation}\label{3.7}
\int \int \frac{d^nk\,d^nq}{(2 \pi)^{(2n)}}\,\frac{2}{D^3}=\frac{\mu^{8-2n} \Gamma(3-n)}{ (\alpha \beta)^{n/2} (4 \pi)^n}\left(x_{i}\mu_{i}^2-i \epsilon-\gamma \frac{}{} p^2\right)^{n-3}
\end{equation}
 Now we define $4-n=\epsilon$ and then take the limit of $n\rightarrow4$ in (\ref{3.7}). As a result the  momentum integrals turns out to be as:
 \begin{equation}
 \begin{split}
  \int \int \frac{d^4k\,d^4q}{(2 \pi)^8}\,\frac{2}{D^3}&:=f(x,y,z;p^2,\mu_{i}^2)\\
  &=\,\frac{x_{i}\mu_{i}^2-i \epsilon_{f}-\gamma p^2}{(4\pi)^4 (\alpha \beta)^2}\,\lim_{\epsilon\rightarrow 0}\left(-\frac{1}{\epsilon}+\gamma_{E}-\log\left(\frac{4 \pi \,\sqrt{\alpha \beta}\,\mu^2}{x_{i}\mu_{i}^2-i \epsilon-\gamma p^2}\right)+O(\epsilon)\right).
\end{split}
 \end{equation}
 Therefore, the self-energy equation is given by:
 \begin{equation}\label{5.6}
\Sig(p^2)=\frac{g^2}{6}  \prod_{i=1}^{3}\int_{0}^{\infty} d\mu_{i}^2 \rh({\mu_{i}^2}) \int_{0}^{1}\int_{0}^{1}\int_{0}^{1} dx dy dz \delta(x+y+z-1) f(x,y,z;p^2,\mu_{i}^2).
 \end{equation}
To make it useful for the numerical computations, it is needed to get rid off the delta function in the integrand. To proceed we change the $(x,y)$ coordinates to $(\xi, \omega)$ through the following relations:
\begin{equation}
x=\xi\,\omega,\,\,\,\,\,\,\,\,\,\,y=(1-\xi)\,\omega
\end{equation}
It is simple to show that under the above transformations, equation (\ref{5.6}) takes the following form: 
 \begin{equation}\label{5.8}
\Sig(p^2)=\frac{g^2}{6}  \prod_{i=1}^{3}\int_{0}^{\infty} d\mu_{i}^2 \rh({\mu_{i}^2}) \int_{0}^{1}\int_{0}^{1}\omega\, d\omega\, d\xi \,\, f\left(\xi\omega,(1-\xi)\frac{}{}\omega,1-\omega;p^2,\mu_{i}^2\right).
 \end{equation}
\subsection{Pad\'{e} Approximants of $f(x)$ around $x=0$}
\label{AppPade}
In the following, we have rewritten a number of Pad\'{e} approximants of function $f(x)$ around point $x=0$.
As it has already been indicated in the text, $m$ and $n$ in $[m/n]_{f}$ denote the degree of polynomials in numerator and denominator respectively.
\begin{equation}
\begin{split}
 [1/0]_{f}(x)&=f(0)+xf'(0) \\
 [0/1]_{f}(x)&=\frac{f(0)}{1-x \,\frac{f'(0)}{f(0)}}  \\
 [0/2]_{f}(x)&= \frac{ f(0)}{1- x \, \frac{f'(0)}{f(0)}+x^2\,\frac{2 f'(0)^2-f(0) f''(0))}{2 f(0)^2}}  \\
[1/1]_{f}(x)&=\frac{f(0)+x\,\frac{2 f'(0)^2-f(0) f''(0)}{2 f'(0)}}{1- x\,\frac{f''(0)}{2 f'(0)}} \\
[1/2]_{f}(x)&=\frac{f(0)+x\, \frac{6f'(0)^3-6f(0)f'(0)f''(0)+f(0)^2 f'''(0)}{6f'(0)^2-f(0)f''(0)}}{1+x\,\frac{-3 f'(0)f''(0)+f(0)f'''(0)}{6f'(0)^2-3 f(0)f''(0)}+x^2\,\frac{3 f''(0)^2-2f'(0)f'''(0)}{12 f'(0)^2-6f(0)f''(0)}}
    \end{split}
\end{equation}
As it can be seen, to construct $[m/n]_{f}$, we need the first $m+n$ derivatives of $f$ at $x=0$.

\subsection{Comparison Between Exact Solution and  Pad\'{e} Approximate solution for a Toy Integral Equation}
\label{AppComp}
\begin{figure}
  \centering
 \includegraphics[width=0.55\textwidth]
{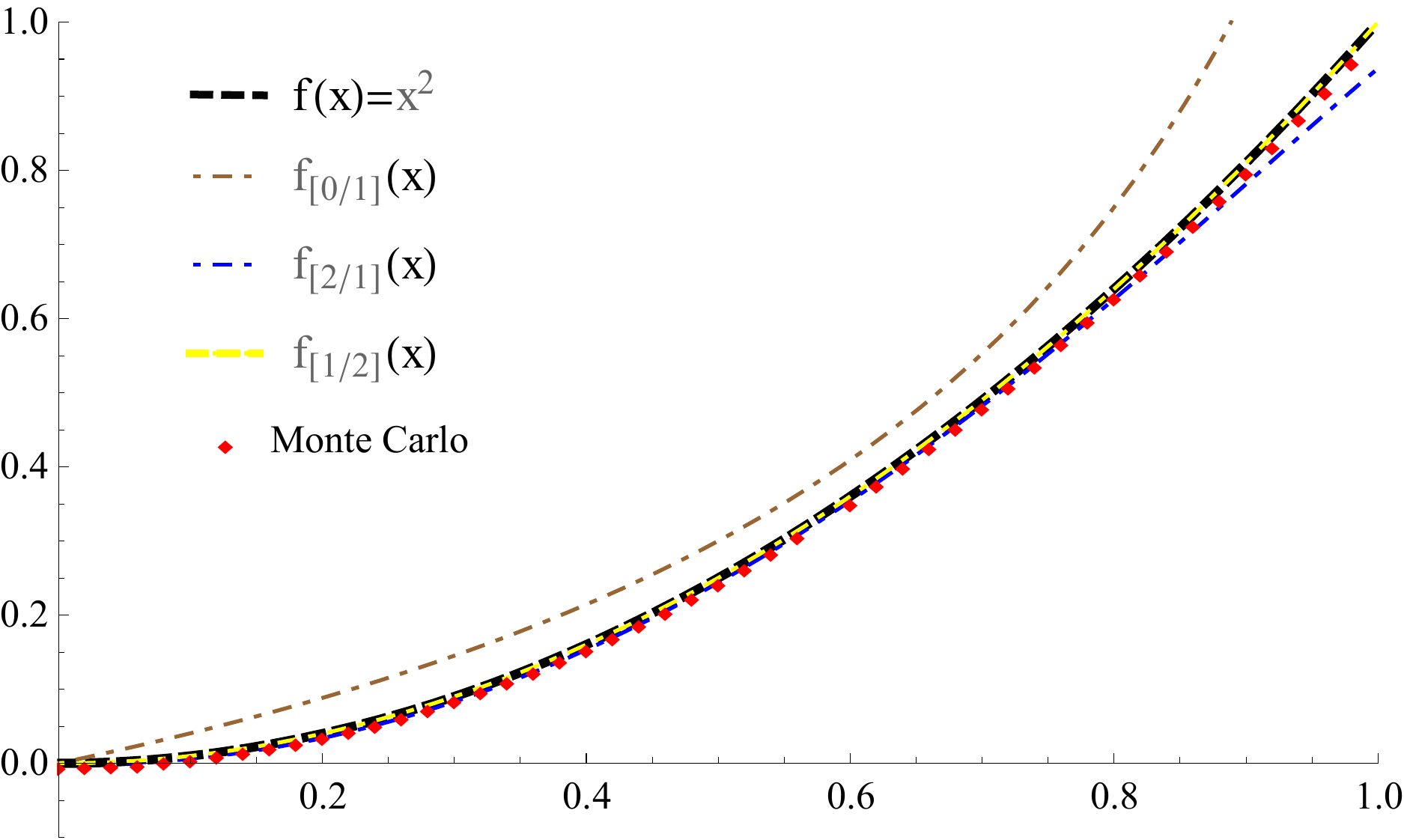}
  \caption{Monte Carlo result versus Pad\'{e} approximate solutions of (\ref{ap4-1}).}\label{toy}
\end{figure}

Let us consider the following toy integral equation:
\begin{equation}\label{ap4-1}
f(x)=\frac{3x}{4}+3\int_{0}^{1}f(y)(x^2-x y)dy.
\end{equation}
It is simple to check that $f(x)=x^2$ exactly solves this equation. Our goal is to use the introduced in section (\ref{VertexPade}) to (\ref{ap4-1}) and   to find its Pad\'{e} approximate solution . Firstly, we must promote $f(x)$ to be also a function of $\epsilon$ as it follows:
\begin{equation}\label{ap4-2}
f(x,\epsilon)=\frac{3x}{4}+3\,\epsilon\int_{0}^{1}f(y,\epsilon)(x^2-x y)dy.
\end{equation}
As we indicated in the text, $[m/n,\epsilon=0]_{f}(x,\epsilon=1)$ is the Pad\'{e} approximate solution
of order $m/n$.\footnote{Just for brevity, however, we write it as $f_{[m/n]}(x)$.  }
 Clearly, for every value of $m+n$, there are $m+n+1$ independent Pad\'{e} approximate solutions.
 In the following we have listed a number of approximate solutions characterized by $m+n\leq4$.
\begin{equation}\label{A.16}
\begin{split}
&1:\,\,\,\,f_{[0/1]}(x)
,\,\,\,\,\,\,\,\,\,\,\,\,\,\,\,\,\,\,\,\,\,\,\,\,\,\,\,\,\,\,\,\,\,\,\,\,\,\,\,\,\,\,\,\,\,\,\,\,\,\,\,\,\,\,\,\,\,\,\,\,\,\,\,\,\,\,\,\,\,\,\,\,\,\,\,\,\,\,\,\,\,\,\,\,\,\,\,\,\,\,\,\,\,\,\,\,\,\,\,\,\,\,\,\,\,\,\,\,\,\,\,\,\,\,\,\,\,\,\,\,\,\,\,\,\,\,f_{[1/0]}(x)=\frac{9}{8}x^2\\
&2:\,\,\,\,f_{[1/1]}(x)=\frac{3x(1-12x+18x^2)}{-28+48x},\,\,\,\,\,\,\,\,\,\,\,\,\,\,\,\,\,\,\,\,\,\,\,\,\,\,\,\,\,\,\,\,\,\,\,\,\,\,\,\,\,\,\,\,\,\,\,\,\,\,\,\,\,\,\,\,\,\,\,\,\,\,\,\,\,\,f_{[2/0]}(x)=\frac{9}{8}x^2-\frac{3}{16}x\\
&3:\,\,\,\,f_{[2/1]}(x)=\frac{3x(1-24x+12x^2)}{-64+48x},\,\,\,\,\,\,\,\,\,\,\,\,\,\,
f_{[1/2]}(x)=x^2,\,\,\,\,\,\,\,\,\,\,\,\,\,\,\,\,f_{[3/0]}=\frac{9}{8}x^2+\frac{3}{8}x\\
&4:\,\,\,\,f_{[1/3]}(x)=x^2,\,\,\,\,\,\,\,\,\,\,\,\,\,\,\,\,\,\,\,\,\,\,\,\,\,\,\,\,\,\,\,\,\,\,\,\,\,\,\,\,\,\,\,\,\,\,\,\,\,\,\,\,\,\,\,\,\,\,\,\,\,\,\,\,\,\,\,\,\,\,\,\,\,\,\,\,\,\,\,\,\,\,\,\,\,\,\,\,\,\,\,\,\,\,\,\,\,\,\,\,\,\,\,\,\,\,\,\,\,\,\,\,\,\,f_{[4/0]}=\frac{9}{8}x^2+\frac{51}{128}x
\end{split}
\end{equation}
Interestingly, very early in the tail of solutions we reach not only to a convergence for successive
approximates, but also we obtain the "
\textbf{exact}" solution of integral equation just at $(m+n=4)^{th}$ level of Pad\'{e} approximation.
The convergence pointed out above may be simply pursued in Fig.(\ref{toy}). In this figure we have also depicted the Monte Carlo solution for (\ref{ap4-1}) obtained via using method developed in \citep{bdmc1}.

Before ending this subsection let us compare the Pad\'{e} approximate solutions with the solutions that might be obtained from Taylor expansion. In fact, the $f_{[m+n/0]}(x)$ solutions in (\ref{A.16}) are nothing but the Taylor approximate solutions which have been found through expanding $f(x,\epsilon)$ around $\epsilon=0$ and then  evaluating at $\epsilon=1$. As it is obvious, although Taylor approximate solutions are converging to the exact solution too, however, for the same value of $m+n$ they are not as well as Pad\'{e} approximants.

\end{document}